\documentclass[aps,prl,twocolumn]{revtex4}
\usepackage{graphicx}
\usepackage{dcolumn}
\usepackage{bm}
\usepackage{amssymb}

\usepackage[]{hyperref}
\usepackage{color} % color

\begin{document}
%%%%%%%%%%%%%%%%%%%%%%%%%%%%%%%%%%%%%%%%%%%%%%%%%%%%%%%%%%%%%%%%%%%%%%%%%%%%%%%%
%%%%%%%%%%%%%%%%%%%%%%%%%%%%%%%%%%%%%%%%%%%%%%%%%%%%%%%%%%%%%%%%%%%%%%%%%%%%%%%%

\title{Large Cross-Phase Modulations at the Few-Photon Level}

\author{Zi-Yu Liu,$^{1}$ Yi-Hsin Chen,$^2$ Yen-Chun Chen,$^{1}$ Hsiang-Yu Lo,$^{1}$ Pin-Ju Tsai,$^1$ Ite A. Yu,$^2$$^\dag$ Ying-Cheng Chen,$^3$ and Yong-Fan Chen,$^1$}

\email{yfchen@mail.ncku.edu.tw} \altaffiliation{$^\dag$ Email address: yu@phys.nthu.edu.tw}

%\altaffiliation{$^\dag$Present address: Department of Electrophysics, National Chiao Tung University, Hsinchu 30013, Taiwan}
%\altaffiliation{$^\ddag$Present address: Institute for Quantum Electronics, ETH Z\"{u}rich, 8093 Z\"{u}rich, Switzerland}

\affiliation{$^1$Department of Physics, National Cheng Kung University, Tainan 70101, Taiwan \\
$^2$Department of Physics and Frontier Research Center on Fundamental and Applied Sciences of Matters, National Tsing Hua
University, Hsinchu 30013, Taiwan \\
$^3$Institute of Atomic and Molecular Sciences, Academia Sinica, Taipei 10617, Taiwan}

\date{\today}

%%%%%%%%%%%%%%%%%%%%%%%%%%%%%%%%%%%%%%%%%%%%%%%%%%%%%%%%%%%%%%%%%%%%%%%%%%%%%%%%
%%%%%%%%%%%%%%%%%%%%%%%%%%%%%%%%%%%%%%%%%%%%%%%%%%%%%%%%%%%%%%%%%%%%%%%%%%%%%%%%

\begin{abstract}

We demonstrate an efficient cross-phase modulation (XPM) based on a closed-loop double-$\Lambda$ system. The property of the double-$\Lambda$
medium can be controlled by changing the phases of the applied optical fields. This phase-dependent XPM scheme can achieve large phase
modulations at low-light intensities without requiring cavities or tightly focusing of laser beams. With this scheme, we observe a $\pi$-level
phase shift with two pulses both consisting of 8 photons in cold rubidium atoms. Such novel scheme provides a simple route to generate strong
interactions between photons and may have potential applications in all-optical quantum signal processing.

\end{abstract}

%%%%%%%%%%%%%%%%%%%%%%%%%%%%%%%%%%%%%%%%%%%%%%%%%%%%%%%%%%%%%%%%%%%%%%%%%%%%%%%%
%%%%%%%%%%%%%%%%%%%%%%%%%%%%%%%%%%%%%%%%%%%%%%%%%%%%%%%%%%%%%%%%%%%%%%%%%%%%%%%%

%03.67.-a Quantum information
%32.80.Qk Coherent control of atomic interactions with photons
%42.25.Hz Interference
%42.50.Gy Effects of atomic coherence on propagation, absorption, and amplification of light; electromagnetically induced transparency and absorption

%%%%%%%%%%%%%%%%%%%%%%%%%%%%%%%%%%%%%%%%%%%%%%%%%%%%%%%%%%%%%%%%%%%%%%%%%%%%%%%%
%%%%%%%%%%%%%%%%%%%%%%%%%%%%%%%%%%%%%%%%%%%%%%%%%%%%%%%%%%%%%%%%%%%%%%%%%%%%%%%%

\maketitle

%%%%%%%%%%%%%%%%%%%%%%%%%%%%%%%%%%%%%%%%%%%%%%%%%%%%%%%%%%%%%%%%%%%%%%%%%%%%%%%%
%%%%%%%%%%%%%%%%%%%%%%%%%%%%%%%%%%%%%%%%%%%%%%%%%%%%%%%%%%%%%%%%%%%%%%%%%%%%%%%%
\newcommand{\FigOne}{
    \begin{figure}[t] %Fig.1
    \includegraphics[width=8.00cm]{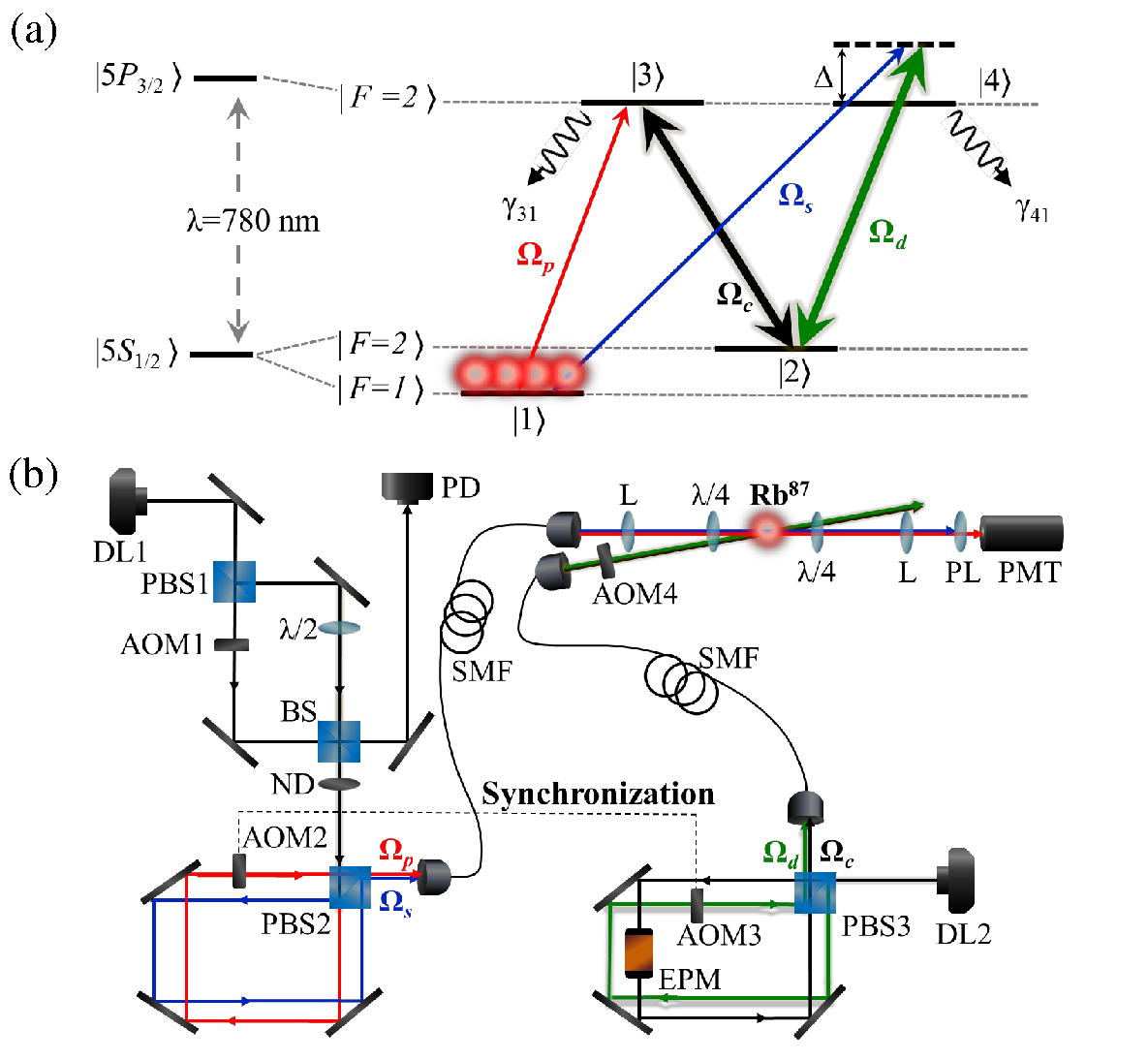}
    \caption{
Energy level scheme and experimental apparatus. (a) Energy levels of $^{87}{\rm Rb}$ $D_2$-line transition for the double-$\Lambda$ experiment.
Signal detuning, $\Delta$, is defined as $\omega_s - \omega_{24}$, where $\omega_s$ and $\omega_{24}$ are the frequencies of the signal field
and the $|2\rangle \leftrightarrow |4\rangle$ transition, respectively. (b) Schematic diagram of the experimental setup. DL, diode laser; PBS,
polarizing-beam splitter; AOM, acousto-optic modulator; $\lambda$/4, quarter-wave plate; $\lambda$/2, half-wave plate; PL, polarizer; ND,
neutral density filter; SMF, single-mode fiber; L, Lens; PD, photo detector; EPM, electro-optic phase modulator; PMT, photomultiplier tube.}
    \label{fig:setup}
    \end{figure}
}
%%%%%%%%%%%%%%%%%%%%%%%%%%%%%%%%%%%%%%%%%%%%%%%%%%%%%%%%%%%%%%%%%%%%%%%%%%%%%%%%
\newcommand{\FigTwo}{
    \begin{figure}[t] %Fig.2
    \includegraphics[width=8.00cm]{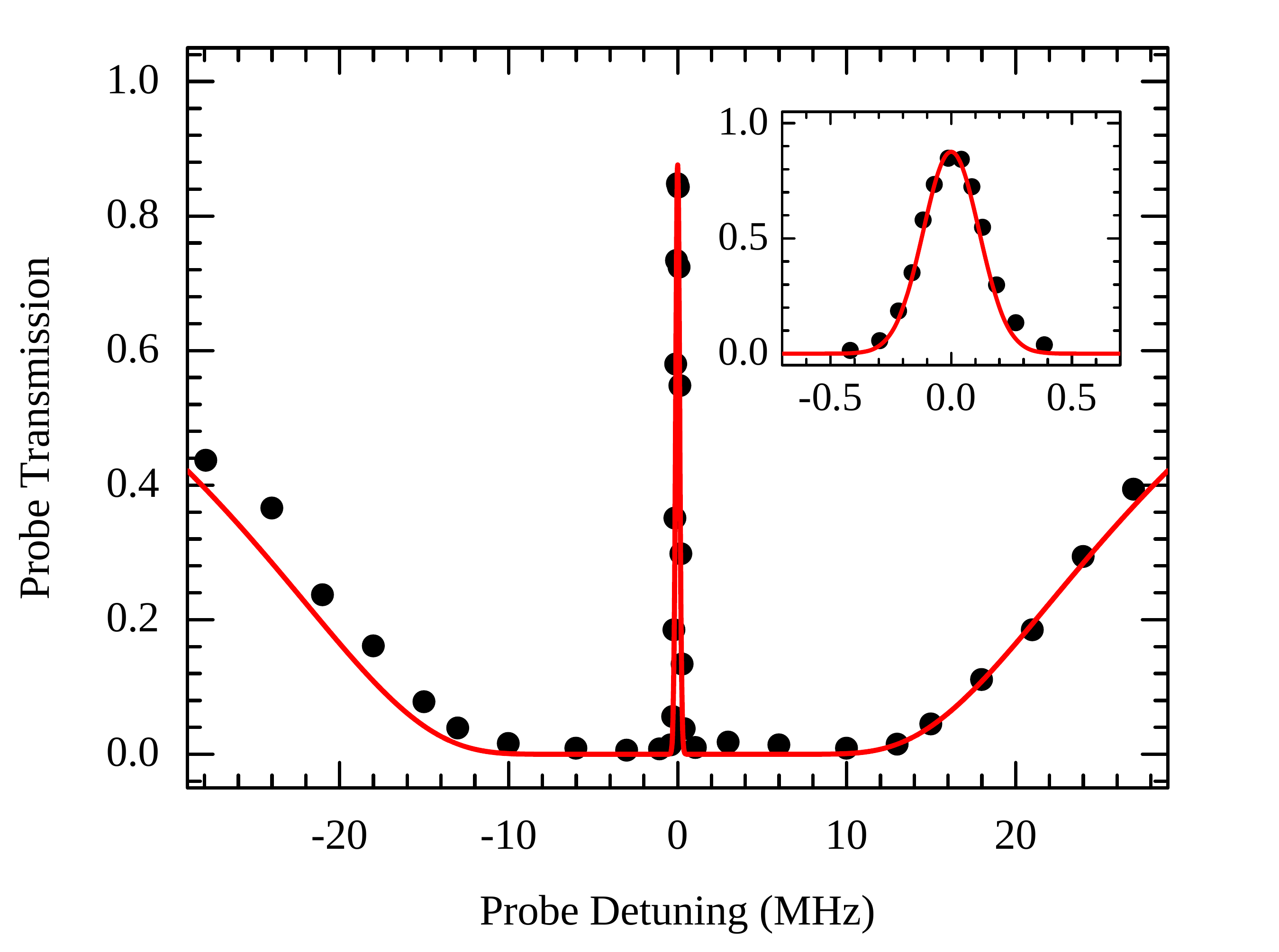}
    \caption{
Observed EIT transmission versus probe field detuning. The black circles and red line represent the measurement data and theoretical curve,
respectively. The inset shows the EIT transmission window. Probe detuning is defined as $\omega_p - \omega_{13}$, where $\omega_p$ and
$\omega_{13}$ are the frequencies of the probe field and the $|1\rangle \leftrightarrow |3\rangle$ transition, respectively. The parameters for
the theoretical curve are $\alpha_{p}=52$, $|\Omega_c|=0.7\Gamma$, $\gamma_{21}=0.001\Gamma$, and $\gamma_{31}=\gamma_{41}=1.25\Gamma$.}
    \label{fig:EIT}
    \end{figure}
}
%%%%%%%%%%%%%%%%%%%%%%%%%%%%%%%%%%%%%%%%%%%%%%%%%%%%%%%%%%%%%%%%%%%%%%%%%%%%%%%%
\newcommand{\FigThree}{
    \begin{figure}[t] %Fig.3
    \includegraphics[width=8.00cm]{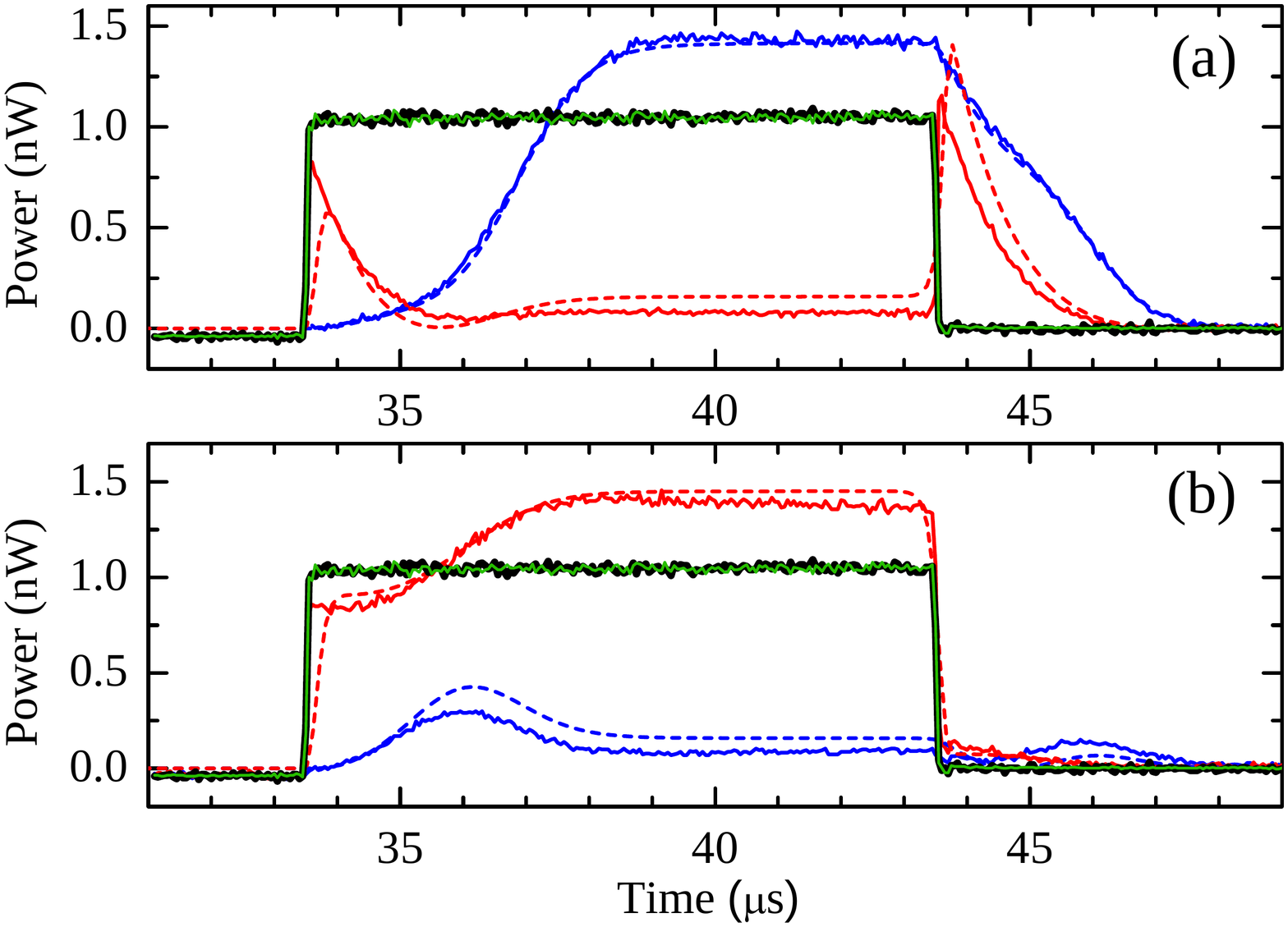}
    \caption{
Phase-dependent double-$\Lambda$ experiment in the pulsed regime. The solid and dashed lines represent the experimental data and theoretical
curves, respectively. The black (green) lines are the input probe (signal) pulses; the blue (red) are the transmitted probe (signal) pulses. The
parameters for the theoretical curves (dashed lines) are $\alpha_{p}=46$, $|\Omega_c| = |\Omega_d| = 0.7\Gamma$, $\Delta = 13\Gamma$,
$\gamma_{21}=0.001\Gamma$, $\gamma_{31}=\gamma_{41}=1.25\Gamma$. (a) the relative phase $\phi_{r}$ = 1.5 rad. (b) $\phi_{r}$ = 4.5 rad. The peak
powers of both the probe and signal pulses are 1 nW, corresponding to around 40,000 photons per pulse.}
    \label{fig:XPMPulse}
    \end{figure}
}
%%%%%%%%%%%%%%%%%%%%%%%%%%%%%%%%%%%%%%%%%%%%%%%%%%%%%%%%%%%%%%%%%%%%%%%%%%%%%%%%
\newcommand{\FigFour}{
    \begin{figure}[t] %Fig.4
    \includegraphics[width=8.00cm]{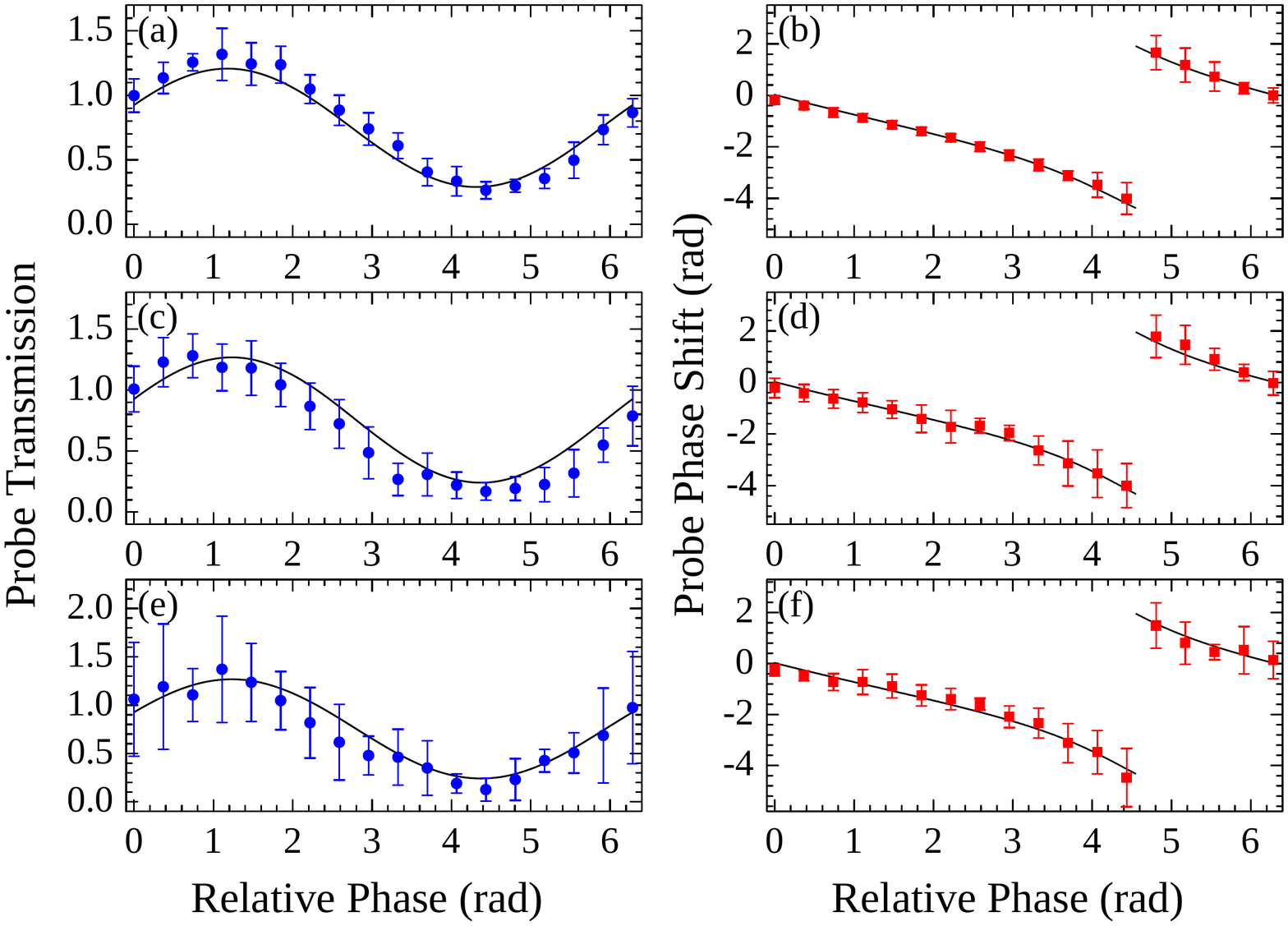}\label{fig:fig4}
    \caption{
Phase modulations at the few-photon level. Dependence of the transmission and the phase shift of the probe pulse on the relative phase
$\phi_{r}$. The numbers of both the probe and the signal photons are around 400 in (a) and (b); 40 in (c) and (d); 8 in (e) and (f). Circles
(squares) represent the transmission (phase shift) of the probe pulse. The blue and red lines are the theoretical curves of the probe
transmission and phase shift, respectively. The parameters for the theoretical curves are $\alpha_{p}=50$, $|\Omega_c| = |\Omega_d| =
0.7\Gamma$, $\Delta = 13\Gamma$, $\gamma_{21}=0.001\Gamma$, $\gamma_{31}=\gamma_{41}=1.25\Gamma$.}
    \label{fig:XPM}
    \end{figure}
}
%%%%%%%%%%%%%%%%%%%%%%%%%%%%%%%%%%%%%%%%%%%%%%%%%%%%%%%%%%%%%%%%%%%%%%%%%%%%%%%%
%%%%%%%%%%%%%%%%%%%%%%%%%%%%%%%%%%%%%%%%%%%%%%%%%%%%%%%%%%%%%%%%%%%%%%%%%%%%%%%%

%\section{Abstract and Introduction}

%The realization of all-optical $\pi$ phase modulation, ultimately at the single-photon level, is a challenging task in quantum information
%science~\cite{TurchetteCavityXPM, FushmanCavityXPM, MatsudaFiberXPM}.

The realization of large cross-phase modulations (XPM) at low-light intensities, ultimately at the single-photon level, is an important but
challenging task in quantum information science~\cite{ZeilingerQI,NielsenQI,YamamotoQND}. To reach this goal, one often requires high-finesse
cavities to enhance nonlinear interactions between photons~\cite{TurchetteCavityXPM,FushmanCavityXPM}. However, cavity-based experiments require
many compromises such as balancing cavity bandwidth and light-matter coupling strength, which remain technical difficulties. Another promising
approach for generating strong photon-photon interaction is electromagnetically induced transparency (EIT)~\cite{HarrisEIT,
LukinEIT,FleischhauerEIT}, but according to the theoretical predictions by Harris and coworkers, the cross-phase shift of the EIT-based Kerr
medium in free space has an upper limit of order 0.1 radians at the single-photon level~\cite{HarrisSLNO}. To date, EIT-based XPM on the order
of micro-radians per photon has been observed in cold atoms ~\cite{LoXPM,SteinbergXPM} and Rb-filled fiber system~\cite{GaetaXPM}. In recent
years, to overcome this upper limit there have been many theoretical proposals and experimental studies on this subject including double
slow-light schemes~\cite{LukinDSLXPM,ShiauDSXPM}, stationary light schemes~\cite{LukinSL,YuSLXPM}, cavity EIT
schemes~\cite{MuckeCavityEIT,ZhuCavityEIT}, or Rydberg EIT schemes~\cite{FleischhauerRydbergXPM, AdamsRydbergEIT, AdamsRydbergBlockade,
LukinRydbergBlockade, DurrRydbergEIT, HofferberthRydbergEIT}. Very recently, two research teams have overcame this upper limit and observed
single-photon cross-phase shifts of $\pi/3$ and $\pi$ by using cavity EIT~\cite{VuleticCavityXPM} and Rydberg EIT~\cite{DurrRydbergXPM},
respectively. This is a great progress toward implementing a photon-photon gate.

%These methods often require a beam tightly focused to a spot size equal to the area of the atomic absorption cross section as well as a medium
%with a large optical depth, which remain technically challenging.

%Although there are several proposed methods to enhance optical nonlinear effects, the interaction between two light fields is normally too small
%for practical applications.

Here we report an experimental observation of a novel XPM scheme based on a phase-dependent double-$\Lambda$ system. With this scheme, we
observe a large cross-phase shift of 3.6$\pm$1.0 radians induced by a light pulse containing around 8 photons in cold rubidium atoms. This XPM
scheme does not require cavities or Rydberg atoms, which provides a simple route to generate strong interactions between photons and obtain
large cross-phase shifts per photon.

%Additionally, we observe a light-amplification phenomenon resulting from coherent light transfer due to the competition between two four-wave
%mixing (FWM) processes in this double-$\Lambda$ system.

%%%%%%%%%%%%%%%%%%%%%%%%%%%%%%%%%%%%%%%%%%%%%%%%%%%%%%%%%%%%%%%%%%%%%%%%%%%%%%%%
%%%%%%%%%%%%%%%%%%%%%%%%%%%%%%%%%%%%%%%%%%%%%%%%%%%%%%%%%%%%%%%%%%%%%%%%%%%%%%%%

%\section{Experimental Details}

In the present study, we investigate a closed-loop double-$\Lambda$ XPM in a laser-cooled $^{87}{\rm Rb}$ atomic system, as depicted in
Fig.~\ref{fig:setup}(a). Cold atomic gas with an optical depth of approximately 50 is produced in a dark spontaneous-force optical trap
(SPOT)~\cite{KetterleSPOT}.A strong coupling field ($\Omega_c$ denotes its Rabi frequency) drives the $|2\rangle \leftrightarrow |3\rangle$
transition to create a transparent medium for a weak probe pulse ($\Omega_p$, $|1\rangle \leftrightarrow |3\rangle$) through quantum
interference. The coupling and probe fields form the first $\Lambda$-type EIT system. The second $\Lambda$-type EIT system is created by a
strong driving field ($\Omega_d$, $|2\rangle \leftrightarrow |4\rangle$) and a weak signal pulse ($\Omega_s$, $|1\rangle \leftrightarrow
|4\rangle$). In the experiment, the coupling and probe fields are right circularly polarized ($\sigma+$) while the driving and signal fields are
left circularly polarized ($\sigma-$). The four laser fields drive the $D_2$-line transition of the $^{87}{\rm Rb}$ atoms to form the
closed-loop double-$\Lambda$ EIT system, as shown in Fig.~\ref{fig:setup}(a).

%The dark SPOT is implemented using a typical magneto-optical trap with a bright capturing region, and two perpendicular repumping beams
%(diameter 2.5 cm, power 0.4 mW) with 5 mm diameter dark areas that drive the $|1\rangle \leftrightarrow |3\rangle$ transition resonantly to form
%a dark region in the center of the trap. The temperature of the cold $^{87}{\rm Rb}$ atoms produced in the dark SPOT is around 300 $\mu$K.

A schematic diagram of the experimental setup is shown in Fig.~\ref{fig:setup}(b). The probe and signal fields are produced using a single diode
laser (DL1); the coupling and driving fields are produced using another diode laser (DL2). DL2 is directly injection locked using an external
cavity diode laser (ECDL, TOPTICA DL 100) with a laser linewidth of around 1 MHz. One beam from the ECDL is sent through a 6.8-GHz electro-optic
modulator (EOM, New Focus 4851). DL1 is injection locked by an intermediate laser seeded with the high-frequency sideband of the EOM output. The
above arrangement is capable of completely eliminating the influence of the carrier of the EOM output on DL1. The probe beam is overlapped with
the signal beam on a polarization beam splitter (PBS2), and then sent to a single-mode fiber (SMF) to obtain the optimal spatial mode-matching.
The $e^{-2}$ diameters of the probe (signal) and coupling (driving) beams are 0.2 mm and 3 mm, respectively. These two beams propagate at an
angle of around $1^\circ$. All of the laser fields are switched on and off via acousto-optic modulators (AOMs). We utilize AOM1 to control the
widths of the probe and signal pulses. The coupling and driving fields are switched on and off via AOM4 [see Fig.~\ref{fig:setup}(b)]. The
experimental data are detected by a photomultiplier tube module (PMT, Hamamatsu H6780-20 and C9663) with a conversion gain of around $9 \times
10^{7}$ V/W, and then recorded using an oscilloscope (Agilent MSO6034A) throughout the experiment. The number of photons of the few-photon
pulses (probe and signal pulses) are also checked by a single-photon counting module (SPCM, Perkin-Elmer SPCM-AQR-13).

When conducting the phase-dependent double-$\Lambda$ experiment, an electro-optic phase modulator (EPM, Thorlabs EO-PM-NR-C1) is applied to vary
the phase of the coupling field ($\Omega_c$). Furthermore, to stabilize the relative phase of the four laser fields, two main setups are
utilized in this experiment. (i) The optical paths of the probe and signal (coupling and driving) fields are arranged in the configuration of a
Sagnac-like interferometer to reduce the path fluctuations between these two beams, as shown in Fig.~\ref{fig:setup}(b). (ii) AOM2 and AOM3 are
driven by the same RF generator through an RF power splitter (Mini-Circuits ZMSC-2-1+).

We utilize a sensitive beat-note interferometer to measure the cross-phase shift of the weak probe pulse. The probe beam is first split into the
transmitted and reflected beams by PBS1 in order to establish the beat-note interferometer [see Fig.~\ref{fig:setup}(b)]. The transmitted beam
passes through the AOM1, which has a driving frequency of 80 MHz, to generate a first-order beam for the probe pulse and then recombines with
the reflected beam from the PBS1 on a beam splitter (BS). One beam from the BS is called the reference beat notes, which is directly received by
a photo detector (PD, New Focus 1801). The other beam, corresponding to the probe beat notes, is detected by a PMT after propagating through the
double-$\Lambda$ medium. The phase shift of the probe pulse is measured by directly comparing the reference and probe beat notes. In this
experiment, only the phase shift within 1 $\mu$s of the end of the probe pulse is measured in order to acquire the steady-state results. The
probe transmission is simultaneously obtained from the amplitude of the probe beat notes. The experimental setup and details of the beat-note
interferometer can be found in Ref.~\cite{LoBeatNote}.

%%%%%%%%%%%%
\FigOne
%%%%%%%%%%%%

%The experiment is run with a repetition rate of 100 Hz, and all the laser fields were switched on or off by AOMs. The timing sequence in the
%experiment is described below. The magnetic field of the magneto-optical trap (MOT) is first switched off. After a 1.4-ms delay, the repumping
%laser of the MOT is switched off, the coupling field is switched on, and then the trapping beams of the MOT are turned off to ensure that the
%entire population is optically pumped to the ground state $|1\rangle$. When all the fields of the MOT are turned off, the probe and the signal
%square pulses are turned on simultaneously.

%However, due to temperature fluctuations, there was a long-term phase drift of approximately 1 radian per hour in the current experiment.%

%%%%%%%%%%%%%%%%%%%%%%%%%%%%%%%%%%%%%%%%%%%%%%%%%%%%%%%%%%%%%%%%%%%%%%%%%%%%%%%%
%%%%%%%%%%%%%%%%%%%%%%%%%%%%%%%%%%%%%%%%%%%%%%%%%%%%%%%%%%%%%%%%%%%%%%%%%%%%%%%%

%\section{Theoretical Model}

To theoretically analyze the behavior of the probe and signal pulses propagating in the double-$\Lambda$ EIT medium, we use the
Maxwell-Schr\"{o}dinger equations below:
\begin{eqnarray}
\frac{\partial\Omega_{p}}{\partial z} + \frac{1}{c}\frac{\partial\Omega_{p}}{\partial t} &= i \frac{\alpha_{p}\gamma_{31}}{2L} \rho_{31},
\label{Eq:MSEprobe}
\end{eqnarray}
\begin{eqnarray}
\frac{\partial\Omega_{s}}{\partial z} + \frac{1}{c}\frac{\partial\Omega_{s}}{\partial t} &= i \frac{\alpha_{s}\gamma_{41}}{2L} \rho_{41},
\label{Eq:MSEsignal}
\end{eqnarray}
where $\Omega_p = |\Omega_p| e^{i\phi_{p}}$ and $\Omega_s = |\Omega_s| e^{i\phi_{s}}$ are the Rabi frequencies of the probe and signal pulses,
respectively. $\phi_{p}$ ($\phi_{s}$) describes the phase information carried by the probe (signal) pulse. $\rho_{31}$ ($\rho_{41}$) is the
slowly-varying amplitude of the optical coherence of the probe (signal) transition. $\alpha_{p} = n \sigma_{13}L$ ($\alpha_{s} = n\sigma_{14}L$)
represents the optical depth of the probe (signal) transition, where $n$ is the number density of the atoms, $\sigma_{13}$($\sigma_{14}$) is the
atomic absorption cross section of the probe (signal) transition, and $L$ is the optical path length of the medium. $\gamma_{31}$ and
$\gamma_{41}$ represent the total coherence decay rates from the $|3\rangle$ and $|4\rangle$ excited states, respectively. We note that the
optical depths of the probe and signal transitions in this experiment are the same ($\alpha_{p}=\alpha_{s}$) because $\sigma_{13}$ is equal to
$\sigma_{14}$ by considering three degenerate Zeeman sublevels, as shown in Fig.~\ref{fig:setup}(a).

In the case where the probe and signal fields are very weak (i.e., $\rho_{11} \simeq 1$), the optical Bloch equations of the slowly-varying
amplitudes of the density-matrix elements are given by:
\begin{eqnarray}
\frac{d}{dt}\rho_{41} &= \frac{i}{2}\Omega_{s} + \frac{i}{2}\Omega_{d}\rho_{21} + \left(i\Delta - \frac{\gamma_{41}}{2}\right)\rho_{41},
\label{Eq:OBEp41}
\end{eqnarray}
\begin{eqnarray}
\frac{d}{dt}\rho_{31} &= \frac{i}{2}\Omega_{p} + \frac{i}{2}\Omega_{c}\rho_{21} - \frac{\gamma_{31}}{2}\rho_{31},
\label{Eq:OBEp31}
\end{eqnarray}
\begin{eqnarray}
\frac{d}{dt}\rho_{21} &= \frac{i}{2}\Omega^{\ast}_{c}\rho_{31} + \frac{i}{2}\Omega^{\ast}_{d}\rho_{41} - \frac{\gamma_{21}}{2}\rho_{21},
\label{Eq:OBEp21}
\end{eqnarray}
where $\Omega_c = |\Omega_c| e^{i\phi_{c}}$ and $\Omega_d = |\Omega_d| e^{i\phi_{d}}$ are the Rabi frequencies of the coupling and driving
transitions, respectively. $\phi_{c}$ ($\phi_{d}$) describes the phase information carried by the coupling (driving) field. $\Delta$ denotes the
detuning of the signal transition [see Fig.~\ref{fig:setup}(a)]. $\gamma_{21}$ represents the dephasing rate of the $|1\rangle$ and $|2\rangle$
ground states. Each parameter in the theoretical model is individually determined from additional experiments as follows: $|\Omega_c|$ is
determined from the separation of the two absorption peaks in the EIT spectrum. $|\Omega_d|$ is determined from the EIT-based photon-switching
effect~\cite{ChenPS}. $\alpha_{p}$ is derived from the delay time of the slow light pulse~\cite{HauSlowLt}. $\gamma_{21}$ is $0.0010(2)\Gamma$,
as estimated by the degree of EIT transparency. $\Gamma = 2\pi \times 6$ MHz is the spontaneous decay rate of the excited states. $\gamma_{31}$
and $\gamma_{41}$ are both $1.25(2)\Gamma$, contributed mostly by the spontaneous decay rate and laser linewidth, as obtained from the spectral
width of the one-photon absorption. Assume $\gamma_{21} = 0$, $\gamma_{31}$ = $\gamma_{41}$ and $\alpha_{p} = \alpha_{s} = \alpha$, the
steady-state solutions of Eqs.~(\ref{Eq:MSEprobe})--(\ref{Eq:OBEp21}) for the probe and signal fields are:
\begin{eqnarray}
    \text{$\Omega_p(\alpha)$}=\frac{1}{\left|\Omega\right|^2}\left[\left|\text{$\Omega_c $}\right|^2\text{$\Omega_p$}(0)+\text{$\Omega_c$} \text{$\Omega_d^*$} \text{$\Omega_s$}(0)\right] \nonumber \\
    +\frac{1}{\left|\Omega\right|^2}\left[\left|\text{$\Omega_d$}\right|^2\text{$\Omega_p$}(0)-\text{$\Omega_c$} \text{$\Omega_d^*$} \text{$\Omega_s$}(0)\right]e^{-i\frac{\alpha }{2\xi }},
\label{Eq:probe}\\
    \text{$\Omega_s(\alpha)$}=\frac{1}{\left|\Omega\right|^2}\left[\left|\text{$\Omega_d $}\right|^2\text{$\Omega_s$}(0)+\text{$\Omega_d$} \text{$\Omega_c^*$} \text{$\Omega_p$}(0)\right] \nonumber
    \\
    +\frac{1}{\left|\Omega\right|^2}\left[\left|\text{$\Omega_c$}\right|^2\text{$\Omega_s$}(0)-\text{$\Omega_d$} \text{$\Omega_c^*$} \text{$\Omega_p$}(0)\right]e^{-i\frac{\alpha }{2\xi }},
\label{Eq:signal}
\end{eqnarray}
where $|\Omega|^{2}=|\Omega_c|^{2}+|\Omega_d|^{2}$, $\xi=i+2\frac{|\Omega_c|^{2}\Delta}{|\Omega|^{2}\gamma_{31}}$. The terms $\Omega_p(0)$ and
$\Omega_s(0)$ represent the incident probe and signal fields, respectively. Under the conditions of $|\Omega_c|=|\Omega_d|$ and
$|\Omega_p(0)|=|\Omega_s(0)|$, we obtain a simple steady-state solutions for the probe and signal fields as follows:
\begin{eqnarray}
&&    \frac{\text{$\Omega_p(\alpha)$}}{\text{$\Omega_p(0)$}}=
    \frac{1}{2}\left[1+e^{-i\text{$\phi_r$}}+\left(1-e^{-i\text{$\phi_r$}}\right)e^{-i\frac{\alpha }{2\xi }}\right],
\label{Eq:probesimple}\\
&&    \frac{\text{$\Omega_s(\alpha)$}}{\text{$\Omega_s(0)$}}=
    \frac{1}{2}\left[1+e^{i\text{$\phi_r$}}+\left(1-e^{i\text{$\phi_r$}}\right)e^{-i\frac{\alpha }{2\xi }}\right].
\label{Eq:signalsimple}
\end{eqnarray}
where the relative phase of the four laser fields, $\phi_{r}$, is defined as $\phi_{p}-\phi_{c}+\phi_{d}-\phi_{s}$. According to
Eqs.~(\ref{Eq:probesimple}) and (\ref{Eq:signalsimple}), when $\Delta = 0$ and $\phi_r=0$, the double-$\Lambda$ medium becomes completely
transparent for both the probe and the signal fields. On the other hand, when $\phi_{r}=\pi$, the medium becomes opaque and has maximum
attenuation for both the probe and the signal fields. This phase-dependent double-$\Lambda$ system with $\Delta = 0$ can be applied in
all-optical switching, as previously described~\cite{ZhuDLambdaPS1}. Here we focus on demonstrating large phase modulations at low-light
intensities with this scheme. Of note, the matched propagation of a pair of slow light pulses in the double-$\Lambda$ medium has been
theoretically discussed in Ref.~\cite{DengDLambda}. Recently, the closed-loop double-$\Lambda$ system can be engineered to achieve broadly
tunable light shifts at low-light intensities has also been theoretically studied in Ref.~\cite{ArtoniXPM}.

%In the case where $\Delta=0$, $|\Omega_p(0)|=|\Omega_s(0)|$, and $|\Omega_c|=|\Omega_d|$, according to Eqs.~(6) and (7),
%$\Omega_p(\alpha_{p})=|\Omega_p(0)|e^{i\phi_{p}}$ and $\Omega_s(\alpha_{s})=|\Omega_s(0)|e^{i\phi_{s}}$ when $\phi_{r}=0$, which means the
%double-$\Lambda$ medium becomes completely transparent for both the probe and the signal fields. On the other hand, when $\phi_{r}=\pi$, the
%medium becomes opaque and has maximum attenuation for both the probe and the signal fields.

%\[
%\Omega_p(\alpha_{p}) = \frac{|\Omega_c|}{|\Omega|^{2}}
%{\left[|\Omega_c||\Omega_p(0)|e^{i\phi_{p}}+|\Omega_d||\Omega_s(0)|e^{i\left(\phi_{p}-\phi_{r}\right)}\right]}+
%\]
%\begin{equation}
%\frac{|\Omega_d|}{|\Omega|^{2}}
%{\left[|\Omega_d||\Omega_p(0)|e^{i\phi_{p}}-|\Omega_c||\Omega_s(0)|e^{i\left(\phi_{p}-\phi_{r}\right)}\right]}e^{-i\frac{\alpha_{p}}{2\xi}},
%\end{equation}
%\\
%\[
%\Omega_s(\alpha_{s}) = \frac{|\Omega_d|}{|\Omega|^{2}}
%{\left[|\Omega_d||\Omega_s(0)|e^{i\phi_{s}}+|\Omega_c||\Omega_p(0)|e^{i\left(\phi_{s}+\phi_{r}\right)}\right]}+
%\]
%\begin{equation}
%\frac{|\Omega_c|}{|\Omega|^{2}}
%{\left[|\Omega_c||\Omega_s(0)|e^{i\phi_{s}}-|\Omega_d||\Omega_p(0)|e^{i\left(\phi_{s}+\phi_{r}\right)}\right]}e^{-i\frac{\alpha_{s}}{2\xi}},
%\end{equation}

%in Ref.~\cite{KorsunskyDLambdaT}.
%the influence of the relative phase of the applied laser fields on the property of the double-$\Lambda$ medium has been theoretically discussed
%in Ref.~\cite{KorsunskyDLambdaT}.
%%%%%%%%%%%%
\FigTwo
%%%%%%%%%%%%

%%%%%%%%%%%%%%%%%%%%%%%%%%%%%%%%%%%%%%%%%%%%%%%%%%%%%%%%%%%%%%%%%%%%%%%%%%%%%%%%
%%%%%%%%%%%%%%%%%%%%%%%%%%%%%%%%%%%%%%%%%%%%%%%%%%%%%%%%%%%%%%%%%%%%%%%%%%%%%%%%

% Experimental result & discussion - EIT Spectrum

We first measure the transmission of a probe pulse propagating through a three-level $\Lambda$-type EIT medium. After all of the lasers and
magnetic fields of the dark SPOT are turned off and the coupling field ($\Omega_c$) is switched on for 100 $\mu$s, the 10-$\mu$s probe square
pulse is switched on to perform the measurement. The experiment is conducted at a repetition rate of 100 Hz. The input power of the probe pulse
is set to 1 nW in the EIT experiment. The Rabi frequency of the coupling transition, $|\Omega_c|$, is 0.7$\Gamma$, corresponding to the coupling
laser power of around 0.5 mW. Figure~\ref{fig:EIT} shows the probe transmission as a function of probe detuning. The inset shows the EIT
transmission window. The measurement data (circles) are in good agreement with the theoretical curve (red line). The theoretical curve is
plotted using the EIT theoretical expression in Ref.~\cite{LoBeatNote}.

%%%%%%%%%%%%
\FigThree
%%%%%%%%%%%%

%%%%%%%%%%%%%%%%%%%%%%%%%%%%%%%%%%%%%%%%%%%%%%%%%%%%%%%%%%%%%%%%%%%%%%%%%%%%%%%%
%%%%%%%%%%%%%%%%%%%%%%%%%%%%%%%%%%%%%%%%%%%%%%%%%%%%%%%%%%%%%%%%%%%%%%%%%%%%%%%%

% Experimental result & discussion - Double-\Lambda XPM in the pulsed regime

Next, we perform the double-$\Lambda$ experiment in the pulsed regime. Figure~\ref{fig:XPMPulse} shows a typical experimental data, where
$\alpha_{p}=46$, $\Delta$ = 13$\Gamma$, $|\Omega_c| = |\Omega_d| = 0.7\Gamma$, and the input powers of both the probe and the signal pulses are
set to 1 nW, corresponding to $|\Omega_p(0)|=|\Omega_s(0)|\approx 0.016\Gamma$ (i.e., $|\Omega_{p(s)}(0)| \ll |\Omega_{c(d)}|$). Here the widths
of both the probe and signal pulses are set to 10 $\mu$s. We utilized the EPM to vary the relative phase $\phi_{r}$ [see
Fig.~\ref{fig:setup}(b)]. The relative phase $\phi_{r}$ is set to 1.5 radians in Fig.~\ref{fig:XPMPulse}(a) and 4.5 radians in
Fig.~\ref{fig:XPMPulse}(b). The solid and dashed lines represent the experimental data and theoretical curves, respectively. The theoretical
curves are plotted by numerically solving Eqs.~(\ref{Eq:MSEprobe})--(\ref{Eq:OBEp21}). The black (green) lines are the input probe (signal)
pulses, and the blue (red) lines are the transmitted probe (signal) pulses. The group-velocity mismatch of the transmitted probe and signal
pulses in Fig.~\ref{fig:XPMPulse} is due to $\Delta \neq 0$. The experimental data also show that the power of the transmitted light exceeds its
input power in the double-$\Lambda$ system. This light-amplification phenomenon is caused by of the coherent light transfer between two N-type
FWM processes ($|1\rangle \rightarrow |3\rangle \rightarrow |2\rangle \rightarrow |4\rangle \rightarrow |1\rangle$ and $|1\rangle \rightarrow
|4\rangle \rightarrow |2\rangle \rightarrow |3\rangle \rightarrow |1\rangle$)~\cite{ChenFWM}. For more detailed discussions for the coherent
light amplification can be found in the Supplemental Material~\cite{SM}.

%It is worth noting that recently the techniques of FWM-based coherent photon conversion in photonic crystal fibres have been proposed to
%implement an efficient quantum computing~\cite{LangfordCPC}. Furthermore, McGuinness~\textit{et al.} have demonstrated that FWM-based quantum
%frequency conversion can preserve the number statics of single-photon states~\cite{McGuinnessQFC,RaymerCPC}.

%%%%%%%%%%%%%%%%%%%%%%%%%%%%%%%%%%%%%%%%%%%%%%%%%%%%%%%%%%%%%%%%%%%%%%%%%%%%%%%%
%%%%%%%%%%%%%%%%%%%%%%%%%%%%%%%%%%%%%%%%%%%%%%%%%%%%%%%%%%%%%%%%%%%%%%%%%%%%%%%%

% Experimental result & discussion - Few-photon all-optical \pi phase modulation

Figure~\ref{fig:XPM} shows the experimental data of the double-$\Lambda$-based XPM at low-light levels. The experimental parameters are the same
as those in Fig.~\ref{fig:XPMPulse} except for the optical depth ($\alpha_{p}=50$). We first perform the double-$\Lambda$ experiment where the
input powers of both the probe and the signal pulses are set to 10 pW, corresponding to around 400 photons per pulse. Figure~\ref{fig:XPM}(a)
and \ref{fig:XPM}(b) show the experimental data of the dependence of the probe transmission and phase shift on the relative phase $\phi_{r}$,
respectively, which are in agreement with the theoretical curves. We subsequently perform the double-$\Lambda$ experiment at the few-photon
level. The input powers of both the probe and the signal pulses in Figs.~\ref{fig:XPM}(c) and \ref{fig:XPM}(e) are reduced to aroun 1 and 0.2
pW, corresponding to around 40 and 8 photons, respectively. Circles (squares) represent the experimental data of the probe transmission (phase
shift). The blue (red) lines are the theoretical curves of the probe transmission (phase shift). Throughout the experiment, the statistical
error bar is evaluated using 6 samples. Each sample is averaged 4096, 16384 and 32768 times for the measurement with 400, 40, and 8 incident
photons, respectively. All error bars in this paper represent a statistical uncertainty of one standard deviation. We note that under the
conditions of $|\Omega_c|=|\Omega_d|$ and $|\Omega_p(0)|=|\Omega_s(0)|$, the probe transmission and phase modulation after propagating through
the double-$\Lambda$ medium are the same according to Eqs.~(\ref{Eq:probesimple}) and (\ref{Eq:signalsimple}). A detailed theoretical analysis
can be found in the Supplemental Material~\cite{SM}.

%a phenomenon of phase jump is observed when the relative phase $\phi_{r}$ is set to around 4.5 radians in the experiment. The behavior of the
%phase jump closely depends on the initial conditions of the double-$\Lambda$ system.

%%%%%%%%%%%%
\FigFour
%%%%%%%%%%%%

As the number of the probe and signal photons decreases, the error bars of the measurement data become large due to smaller signal-to-noise
ratios, as shown in Fig.~\ref{fig:XPM}. Although the data in Fig.~\ref{fig:XPM}(f) possess a large amount of phase noise of around 1 radians,
the measured values are still valid considering the considerable phase shift. For instance, in Fig.~\ref{fig:XPM}(f), a maximum phase shift of
-4.5$\pm$0.9 radians is measured when the relative phase $\phi_{r}$ is set to 4.4 radians. When the signal pulse is absent, we also measure the
probe phase shift of -0.9$\pm$0.1 radians which is consistent with the theoretical predictions. Hence, we conclude that a cross-phase shift of
3.6$\pm$1.0 radians induced by a light pulse containing around 8 photons has been realized with this scheme. So far we do not perform the
experiment using few-photon pulses containing less than 8 photons because the long-term instability of our experimental system prevents us from
improving the signal-to-noise ratios by collecting more data. In principle, this phase-dependent XPM scheme can reach the goal of $\pi$ phase
modulation per photon if one can prepare two phase-coherent single-photon pulses to be as the probe and signal pulses.

In conclusion, we have demonstrated an efficient XPM based on a closed-loop double-$\Lambda$ system. The property of the double-$\Lambda$ medium
can be controlled by changing the phases of the applied optical fields. This phase-dependent XPM scheme can achieve large phase modulations at
low-light intensities without requiring cavities or Rydberg atoms. We have observed a cross-phase shift of 3.6$\pm$1.0 radians induced by a
light pulse containing 8 photons in cold $^{87}{\rm Rb}$ atoms with this scheme. Such novel scheme provides a simple route to generate strong
interaction between photons, and may have potential applications in all-optical quantum signal processing.
%%%%%%%%%%%%%%%%%%%%%%%%%%%%%%%%%%%%%%%%%%%%%%%%%%%%%%%%%%%%%%%%%%%%%%%%%%%%%%%%
%%%%%%%%%%%%%%%%%%%%%%%%%%%%%%%%%%%%%%%%%%%%%%%%%%%%%%%%%%%%%%%%%%%%%%%%%%%%%%%%

%\section{Methods}

%\textbf{Experimental details.}

%%%%%%%%%%%%%%%%%%%%%%%%%%%%%%%%%%%%%%%%%%%%%%%%%%%%%%%%%%%%%%%%%%%%%%%%%%%%%%%%
%%%%%%%%%%%%%%%%%%%%%%%%%%%%%%%%%%%%%%%%%%%%%%%%%%%%%%%%%%%%%%%%%%%%%%%%%%%%%%%%

\section*{ACKNOWLEDGEMENTS}
We acknowledge Hao-Chung Chen, You-Lin Chuang and Ray-Kuang Lee for helpful discussions and Jun-Jie Wu for experimental assistance. This work
was supported by the National Science Council of Taiwan under grants numbers 103-2119-M-006-018 and 104-2119-M-006-002. This work was done under
a collaboration project (Science Vanguard Research Program of MOST) with Ite A. Yu as the project leader and Ying-Cheng Chen and Yong-Fan Chen
as the subproject leaders. Correspondence of the project contents can be addressed to Ite A. Yu; correspondence and requests for material of
this work can be addressed to Yong-Fan Chen. We also acknowledge the support from NCTS of Taiwan.
\end{document}

%%%%%%%%%%%%%%%%%%%%%%%%%%%%%%%%%%%%%%%%%%%%%%%%%%%%%%%%%%%%%%%%%%%%%%%%%%%%%%%%
%%%%%%%%%%%%%%%%%%%%%%%%%%%%%%%%%%%%%%%%%%%%%%%%%%%%%%%%%%%%%%%%%%%%%%%%%%%%%%%%

\title{Supplemental Material for
\\"Large Cross-Phase Modulations at the Few-Photon Level"}

\date{\today}

%%%%%%%%%%%%%%%%%%%%%%%%%%%%%%%%%%%%%%%%%%%%%%%%%%%%%%%%%%%%%%%%%%%%%%%%%%%%%%%%
%%%%%%%%%%%%%%%%%%%%%%%%%%%%%%%%%%%%%%%%%%%%%%%%%%%%%%%%%%%%%%%%%%%%%%%%%%%%%%%%

\maketitle
%%%%%%%%%%%%%%%%%%%%%%%%%%%%%%%%%%%%%%%%%%%%%%%%%%%%%%%%%%%%%%%%%%%%%%%%%%%%%%%%
%%%%%%%%%% Prefix a "S" to all equations, figures, tables and reset the counter %%%%%%%%%%
\setcounter{equation}{0}
\setcounter{figure}{0}
\setcounter{table}{0}
\setcounter{page}{1}
\makeatletter
\renewcommand{\theequation}{S\arabic{equation}}
\renewcommand{\thefigure}{S\arabic{figure}}
\renewcommand{\bibnumfmt}[1]{[S#1]}
\renewcommand{\citenumfont}[1]{S#1}
%%%%%%%%%%%%%%%%%

%%%%%%%%%%%%%%%%%%%%%%%%%%%%%%%%%%%%%%%%%%%%%%%%%%%%%%%%%%%%%%%%%%%%%%%%%%%%%%%%
%%%%%%%%%%%%%%%%%%%%%%%%%%%%%%%%%%%%%%%%%%%%%%%%%%%%%%%%%%%%%%%%%%%%%%%%%%%%%%%%
%\pacs{42.50.Gy, 32.80.Qk, 42.65.-k, 42.50.-p}
%42.50.Gy Effects of atomic coherence on propagation, absorption, and amplification of light; electromagnetically induced transparency and absorption
%32.80.Qk Coherent control of atomic interactions with photons
%42.65.-k Nonlinear optics
%42.50.-p Quantum optics
%03.67.-a Quantum information

%%%%%%%%%%%%%%%%%%%%%%%%%%%%%%%%%%%%%%%%%%%%%%%%%%%%%%%%%%%%%%%%%%%%%%%%%%%%%%%%
%%%%%%%%%%%%%%%%%%%%%%%%%%%%%%%%%%%%%%%%%%%%%%%%%%%%%%%%%%%%%%%%%%%%%%%%%%%%%%%%

\maketitle

%%%%%%%%%%%%%%%%%%%%%%%%%%%%%%%%%%%%%%%%%%%%%%%%%%%%%%%%%%%%%%%%%%%%%%%%%%%%%%%%
%%%%%%%%%%%%%%%%%%%%%%%%%%%%%%%%%%%%%%%%%%%%%%%%%%%%%%%%%%%%%%%%%%%%%%%%%%%%%%%%
\newcommand{\FigOne}{
    \begin{figure}[t] %Fig.1
    \includegraphics[width=8cm]{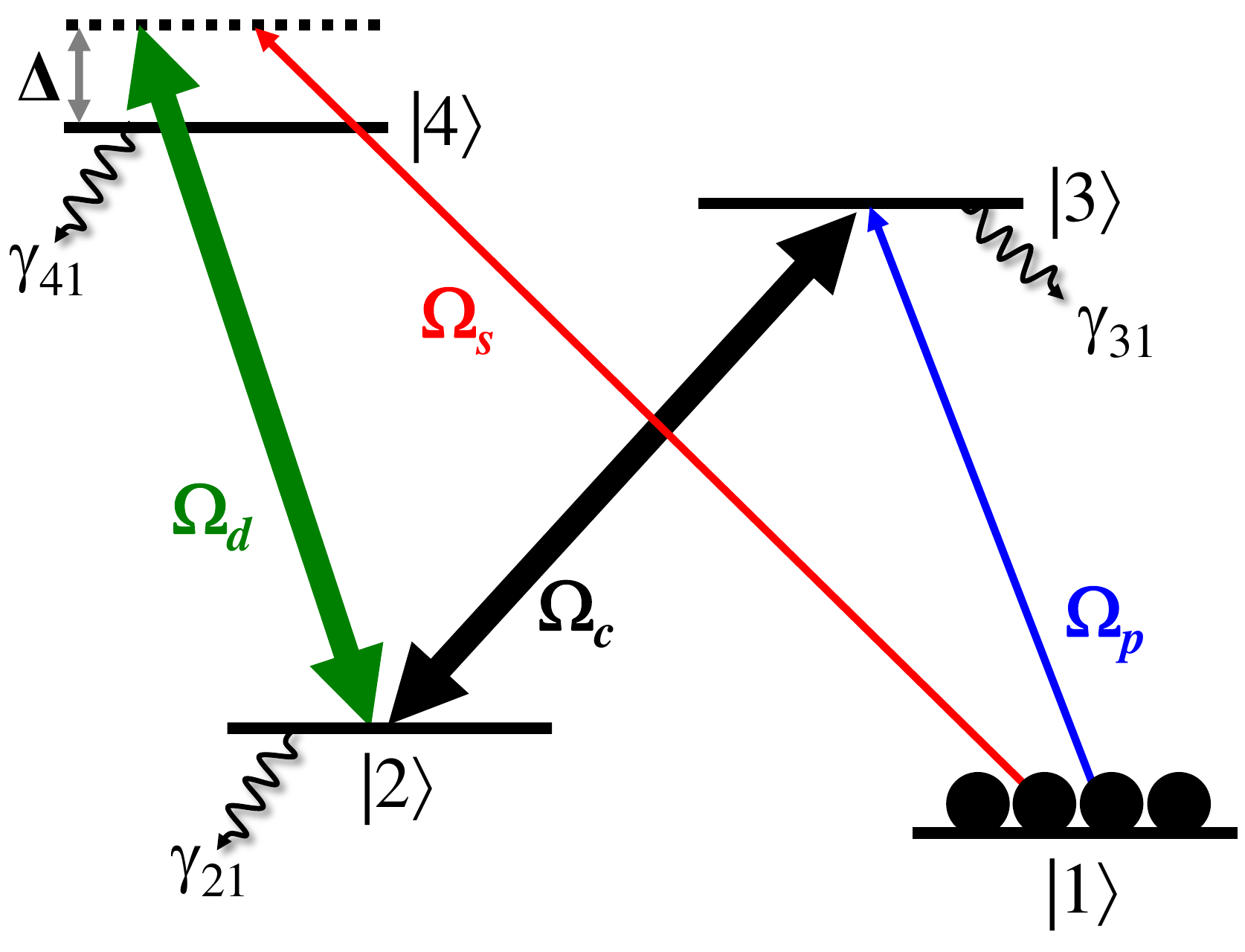}
    \caption{Schematic energy level for a four-level double-$\Lambda$ system. States $|1\rangle$ and $|2\rangle$ are two
    metastable ground states, and states $|3\rangle$ and $|4\rangle$ are two excited states. Weak probe ($\Omega_p$) and strong coupling
    ($\Omega_c$) fields form the first EIT system, and weak signal ($\Omega_s$) and strong driving ($\Omega_d$) fields constitute the second EIT system
    with a detuning of $\Delta$. All the atoms are initially prepared in the state $|1\rangle$. The term $\gamma_{31(41)}$ is the total coherence
    decay rate of the excited state $|3\rangle(|4\rangle)$. The term $\gamma_{21}$ is the dephasing rate of the coherence between the ground states $|1\rangle$ and $|2\rangle$.}
    \label{fig:level}
    \end{figure}
}

%%%%%%%%%%%%%%%%%%%%%%%%%%%%%%%%%%%%%%%%%%%%%%%%%%%%%%%%%%%%%%%%%%%%%%%%%%%%%%%%
\newcommand{\FigTwo}{
    \begin{figure}[t] %Fig.2
    \includegraphics[width=12.5cm]{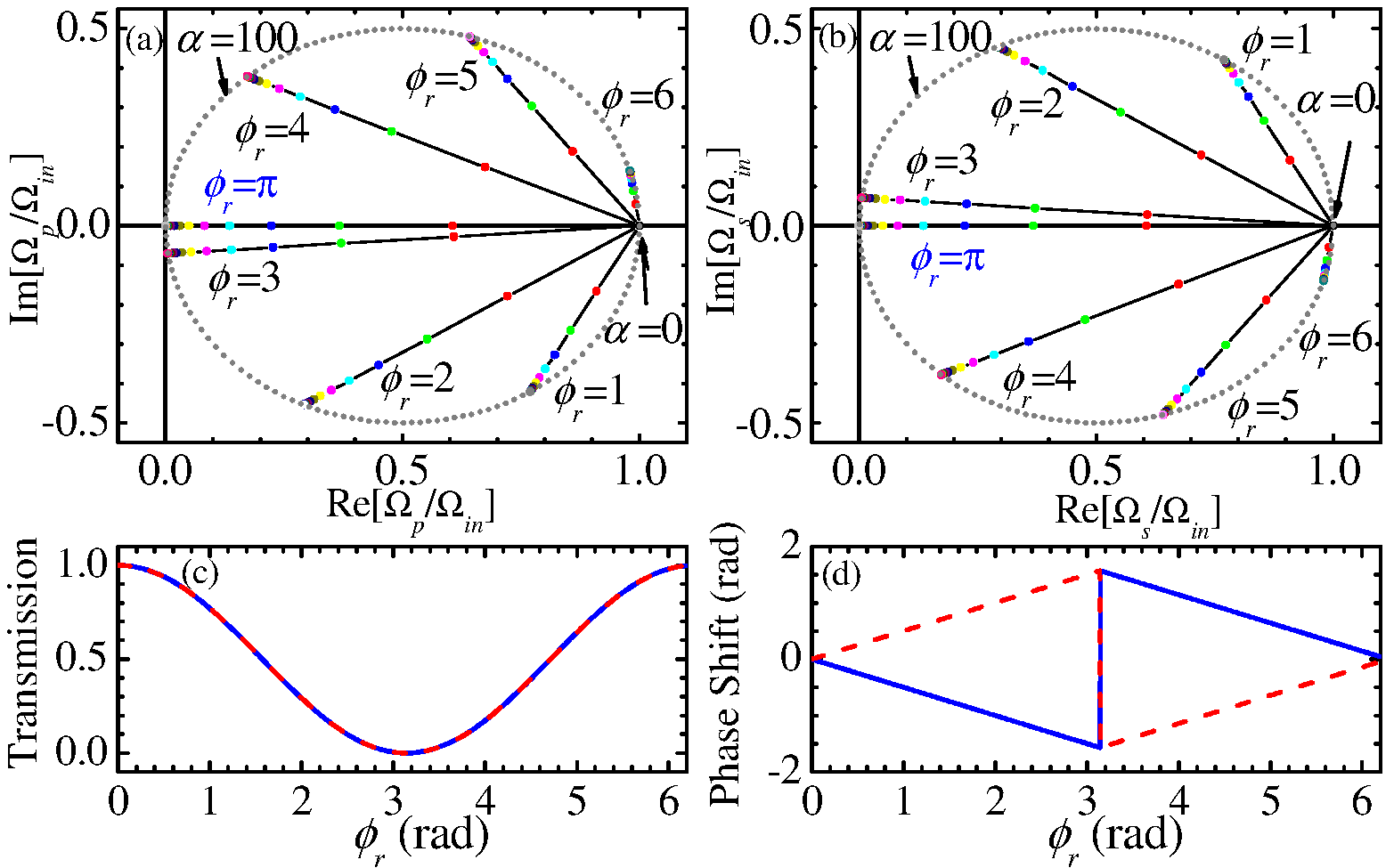}
    \caption{Balanced double-$\Lambda$ EIT system ($\Delta=0$, $|\Omega_c|=|\Omega_d|$, and $|\Omega_p(0)|=|\Omega_s(0)|$). (a) and (b)
    show phase diagrams of the probe and signal fields, respectively, plotted according to Eqs.~(\ref{Eq:probesimple}) and (\ref{Eq:signalsimple}). The
    optical depth, $\alpha$, increased from 0 to 100, clearly illustrating the phase evolutions of the light fields. The relative phases, $\phi_r$, are
    set from 1 to 6 as well as $\pi$ in units of rad. The dotted lines show the loops of the light fields at various $\phi_r$ values and $\alpha=100$.
    (c) and (d) are graphs of the corresponding transmissions and phase shifts of the probe (blue solid lines) and signal fields (red dashed lines) when
    $\alpha=100$.}
    \label{fig:Fig2}
    \end{figure}
}

\newcommand{\FigThree}{
    \begin{figure}[t] %Fig.3
    \includegraphics[width=12.5cm]{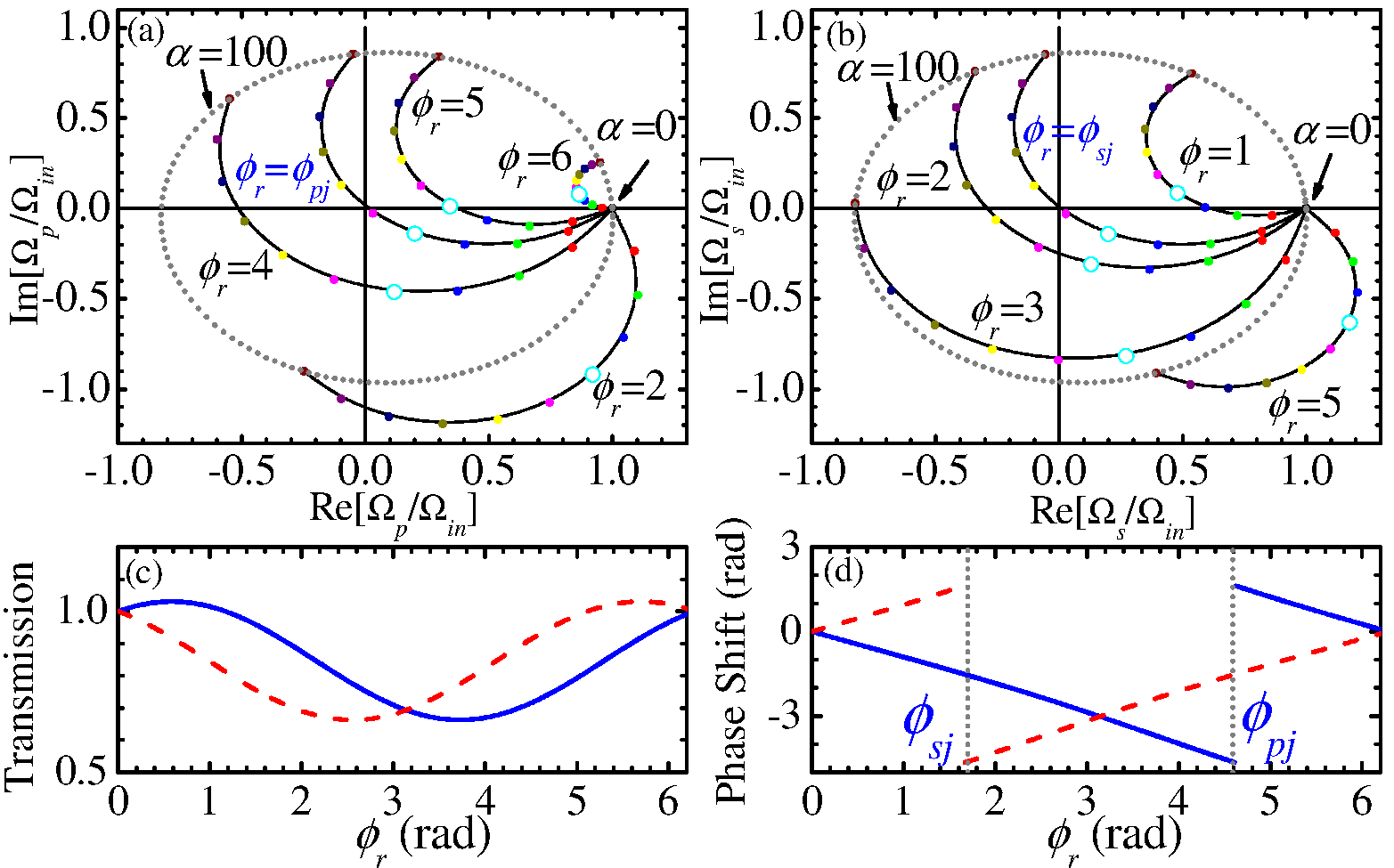}
    \caption{Imbalanced double-$\Lambda$ EIT system ($\Delta \neq 0$, $|\Omega_c|=|\Omega_d|$, and $|\Omega_p(0)|=|\Omega_s(0)|$). (a) and
    (b) show phase diagrams of the probe and signal fields, respectively, plotted according to Eqs.~(\ref{Eq:probesimple}) and (\ref{Eq:signalsimple})
    with $\Delta=16.5 \Gamma$. The values of relative phases, $\phi_r$, are set as shown in the plot and $\phi_{pj}=4.62$ and $\phi_{sj}=1.66$ in units of
    rad. (c) and (d) are graphs of the transmissions and phase shifts of the probe (blue solid lines) and signal fields (red dashed lines) when
    $\alpha=100$.}
    \label{fig:Fig3}
    \end{figure}
}
%%%%%%%%%%%%%%%%%%%%%%%%%%%%%%%%%%%%%%%%%%%%%%%%%%%%%%%%%%%%%%%%%%%%%%%%%%%%%%%%
\newcommand{\FigFour}{
    \begin{figure}[t] %Fig.4
    \includegraphics[width=11cm]{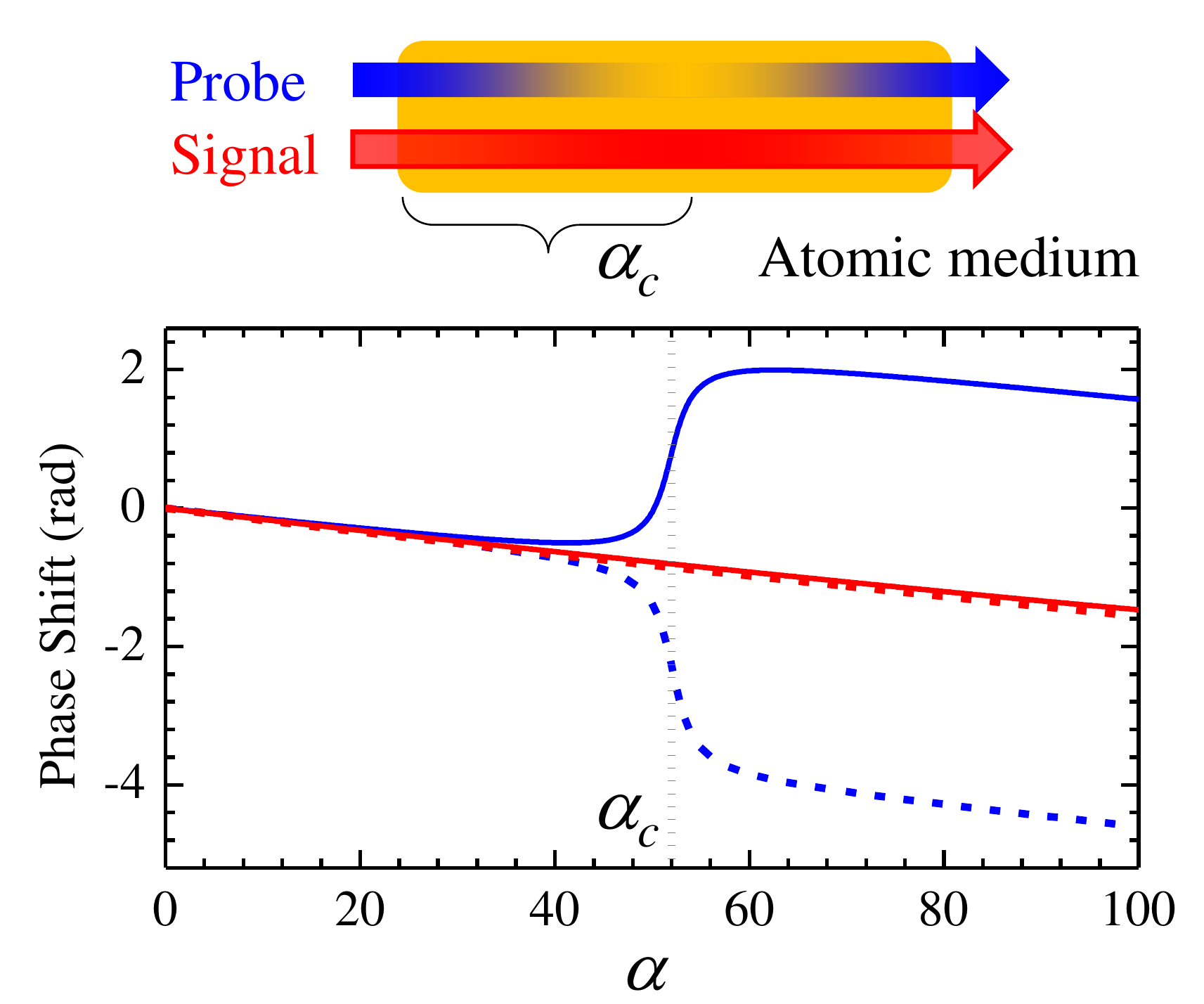}
    \caption{The top illustration shows that the probe field exhausts its energy when the light propagates through a medium with a
    critical optical depth of $\alpha_c$. The probe field is then restored when the light field passes through the remainder of medium. The bottom figure
    shows the phase shifts of the probe (blue) and signal (red) fields with the relative phase right above $\phi_{pj}$ (4.67 rad for the solid lines)
    and below $\phi_{pj}$ (4.57 rad for the dashed lines), plotted according to Eqs.~(\ref{Eq:probesimple}) and (\ref{Eq:signalsimple})
    with $\Delta=16.5 \Gamma$.}
    \label{fig:Fig4}
    \end{figure}
}
%%%%%%%%%%%%%%%%%%%%%%%%%%%%%%%%%%%%%%%%%%%%%%%%%%%%%%%%%%%%%%%%%%%%%%%%%%%%%%%%
\newcommand{\FigFive}{
    \begin{figure}[t] %Fig.5
    \includegraphics[width=12cm]{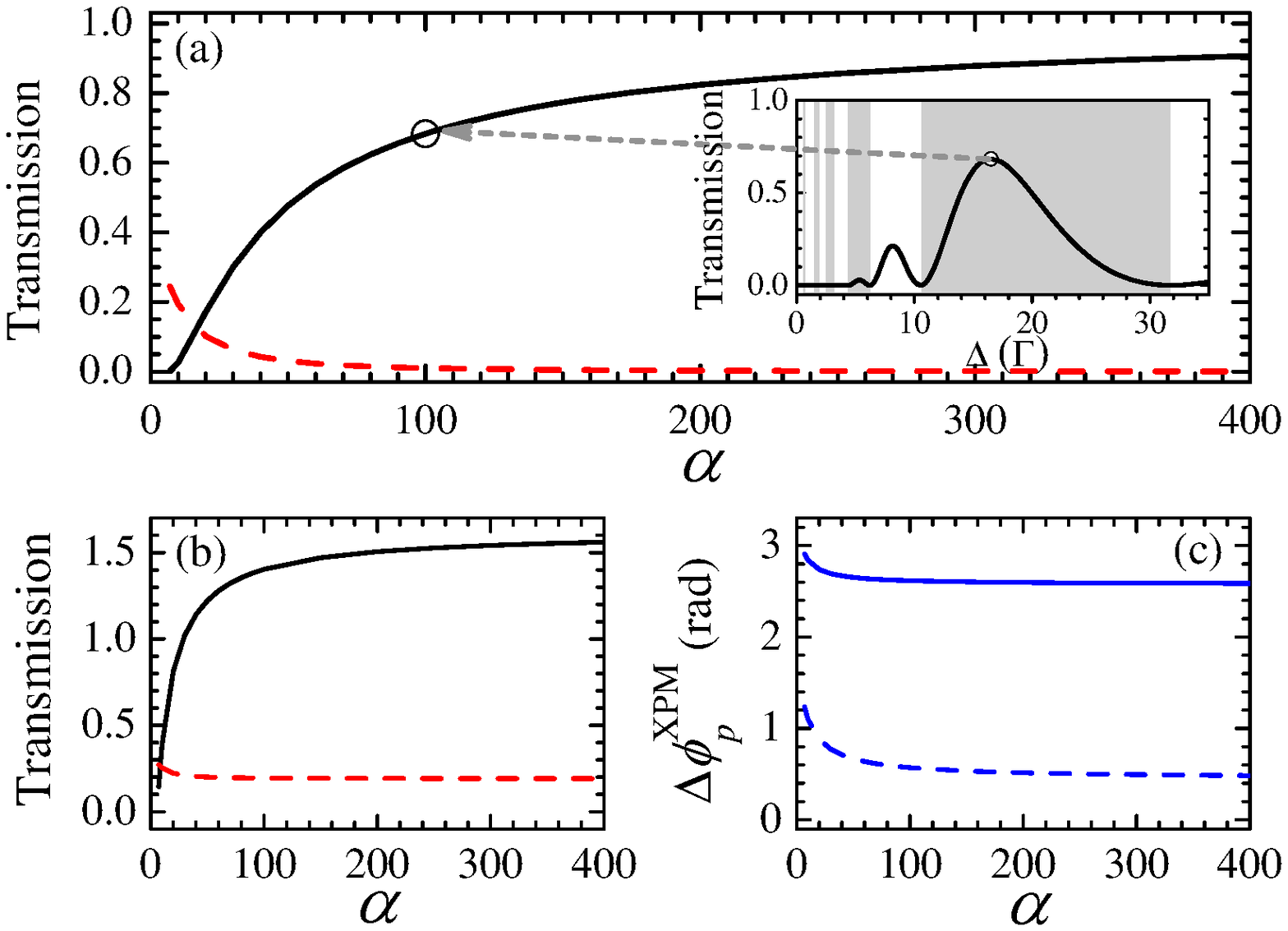}
    \caption{(a) Inset: To achieve a $\pi$ phase shift of the transmitted probe field ($|\Delta \phi_p|$) in a double-$\Lambda$ EIT
    system, the probe field transmission is a function of the detuning $\Delta$ with a fixed optical depth ($\alpha=100$). The gray zones in the figure
    show that the terminal point is located on the negative x-axis. A local maximum of transmission is located at approximately $\Delta=16.5 \Gamma$.
    Main plot: Using various optical depths and the corresponding optimized $\Delta$, we obtain the optimized probe transmission, which is a monotonous
    increasing function (black solid line). Without the signal field (red dashed line), the probe transmission is a monotonous decreasing function of
    $\alpha$ with the corresponding $\Delta$. (b) Simulations similar to those shown in (a) except the phase shift of the transmitted probe field
    is set to $\pi/2$. (c) The phase modulation of the probe field with and without the signal field, $|\Delta\phi_p^{\text{XPM}}|$, as a function of
    $\alpha$. Blue solid and dashed lines represent the phase modulation in the simulations in (a) and (b), respectively.}
    \label{fig:Fig5}
    \end{figure}
}
%%%%%%%%%%%%%%%%%%%%%%%%%%%%%%%%%%%%%%%%%%%%%%%%%%%%%%%%%%%%%%%%%%%%%%%%%%%%%%%%
\newcommand{\FigSix}{
    \begin{figure}[t] %Fig.6
    \includegraphics[width=11cm]{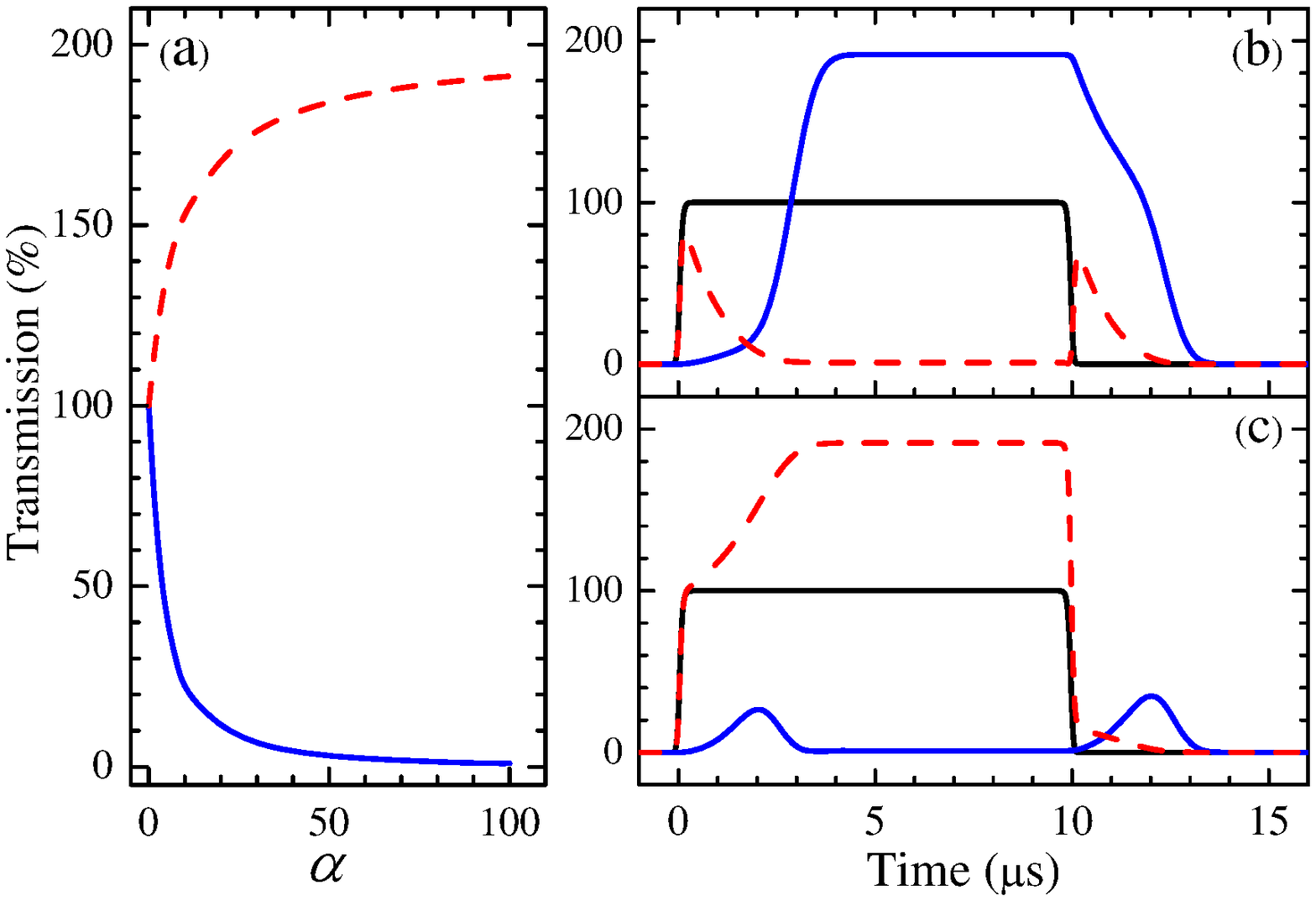}
    \caption{(a) Optimal energy amplification for the signal field versus optical depth $\alpha$ with the optimal detuning $\Delta$ and
    relative phase $\phi_r$ determined using numerical simulation of Eqs.~(\ref{Eq:probesimple}) and (\ref{Eq:signalsimple}). Blue solid and red dashed
    lines represent the transmissions of the probe and signal fields, respectively. An optical depth of 50 (100) enables achieving an amplification
    efficiency of approximately 84$\%$ (91$\%$), as indicated by the red dashed line. (b) and (c) Two slow-light pulses in the double-$\Lambda$ EIT system
    with relative phases $\phi_r$ of 1.53 rad and 4.76 rad to achieve maximal probe (blue solid lines) and signal transmission (red dashed lines) in (b)
    and (c), respectively. In this simulation, $\alpha=100$, $\Delta=34.2\Gamma$, $\Omega_c=\Omega_d=1\Gamma$, and $\gamma_{21}=0$. Black solid lines
    represent two identical incident probe and signal pulses.}
    \label{fig:Fig6}
    \end{figure}
}

%%%%%%%%%%%%%%%%%%%%%%%%%%%%%%%%%%%%%%%%%%%%%%%%%%%%%%%%%%%%%%%%%%%%%%%%%%%%%%%%

%%%%%%%%%%%%%%%%%%%%%%%%%%%%%%%%%%%%%%%%%%%%%%%%%%%%%%%%%%%%%%%%%%%%%%%%%%%%%%%%

\section{Theoretical model} \label{Sec:theory}

We consider an atomic medium consisting of double-$\Lambda$-type four-level transitions with two metastable ground states ($|1\rangle$ and
$|2\rangle$) and two excited states ($|3\rangle$ and $|4\rangle$), as shown in Fig.~\ref{fig:level}. Weak probe (with the Rabi frequency
$\Omega_p$) and strong coupling ($\Omega_c$) fields form the first electromagnetically induced transparency (EIT) system, and weak signal
($\Omega_s$) and strong driving ($\Omega_d$) fields constitute the second EIT system. For an individual EIT system, $\Omega_c$ ($\Omega_d$)
manipulates the transmission of $\Omega_p$ ($\Omega_s$) through an optical dense medium and causes the $\Omega_p$ ($\Omega_s$) to become
transparent because of destructive quantum interference. When the conditions $|\Omega_c| \gg |\Omega_p|$ and $|\Omega_d| \gg |\Omega_s|$ are
satisfied, all the atoms remain in the ground state $|1\rangle$ and the contribution of the probe and signal fields can be treated as a
perturbation in the derivation of the following equations. The double-$\Lambda$ EIT medium can induce two N-type four-wave mixing (FWM)
processes: first, $|1\rangle \rightarrow |3\rangle \rightarrow |2\rangle \rightarrow |4\rangle \rightarrow |1\rangle $, generating the signal
field; and second, $|1\rangle \rightarrow |4\rangle \rightarrow |2\rangle \rightarrow |3\rangle \rightarrow |1\rangle $, generating the probe
field \cite{ChenFWM2014}. Thus, the energy as well as the phases of the probe and signal fields are coherently transferred via these two FWM
paths. We begin deriving equations from the interaction Hamiltonian between atoms and optical fields and the equations of motion of the density
matrix operators. The steady-state analytical solutions for the double-$\Lambda$ EIT system can be obtained by solving the first-order optical
Bloch equations (OBEs) of the density matrix operators and the Maxwell-Schr\"{o}dinger equations (MSEs) of the probe and signal pulses as
follows:

\begin{eqnarray}
    \frac{d}{dt}\rho_{41} =
    \frac{i}{2}\Omega_{s} + \frac{i}{2}\Omega_{d}\rho_{21} + \left(i\Delta -\frac{\gamma_{41}}{2}\right)\rho_{41},
\label{Eq:OBEp41}\\
    \frac{d}{dt}\rho_{31} = \frac{i}{2}\Omega_{p} + \frac{i}{2}\Omega_{c}\rho_{21} -
    \frac{\gamma_{31}}{2}\rho_{31},
\label{Eq:OBEp31} \\
    \frac{d}{dt}\rho_{21} = \frac{i}{2}\Omega^{\ast}_{c}\rho_{31} +
    \frac{i}{2}\Omega^{\ast}_{d}\rho_{41} - \frac{\gamma_{21}}{2}\rho_{21},
\label{Eq:OBEp21}\\
    \frac{\partial\Omega_{p}}{\partial z} + \frac{1}{c}\frac{\partial\Omega_{p}}{\partial t} = i \frac{\alpha_{p}\gamma_{31}}{2L} \rho_{31},
\label{Eq:MSEprobe}\\
    \frac{\partial\Omega_{s}}{\partial z} + \frac{1}{c}\frac{\partial\Omega_{s}}{\partial t} = i \frac{\alpha_{s}\gamma_{41}}{2L} \rho_{41},
\label{Eq:MSEsignal}
\end{eqnarray}
where $\alpha_{p(s)}$ represents the optical depth of the probe (signal) field transition; $\rho_{ij}$ is the slowly varying amplitude of the
coherence between states $|i\rangle$ and $|j\rangle$; $\gamma_{31(41)} \equiv \Gamma_{3(4)}$ + $\gamma_{3(4)}$ is the total coherence decay rate
of the excited state $|3\rangle$ ($|4\rangle$), where $\Gamma_{3(4)}$ and $\gamma_{3(4)}$ represent the total spontaneous decay rate of the
excited state $|3\rangle$ ($|4\rangle$) and the energy-conserving dephasing rate, respectively \cite{EIT2005}; $\gamma_{21}$ is the dephasing
rate of the coherence between the ground states $|1\rangle$ and $|2\rangle$; $L$ is the optical path length of the medium; and $\Delta$ denotes
the detuning of the signal field transition.

%%%%%%%%%%%%%%%%%%%%%%%%%%%%%%%%%%%%%%%%%%%%%%%%%%%%%%%%%%%%%%%%%%%%%%%%%%%%%%%%
\FigOne
%%%%%%%%%%%%%%%%%%%%%%%%%%%%%%%%%%%%%%%%%%%%%%%%%%%%%%%%%%%%%%%%%%%%%%%%%%%%%%%%

For simplicity, we assume $\alpha_p=\alpha_s\equiv\alpha$, $\gamma_{31}=\gamma_{41}\equiv\Gamma$, and $\gamma_{21}=0$. With the time-derivative
terms being zero, we derive the steady-state solutions by solving the first-order OBEs [Eqs.~(\ref{Eq:OBEp41})--(\ref{Eq:OBEp21})] as follows:

\begin{eqnarray}
&&    \text{$\rho_{21}$}=
    \frac{\text{$\Omega_p$} \text{$\Omega^{\ast}_c$} (\text{$2\Delta+i \Gamma$}) + \text{$\Omega_s$} \text{$\Omega^{\ast}_d$} (\text{$i\Gamma$})}{D},\nonumber \\
&&    \text{$\rho_{31}$}=
    \frac{\text{$\Omega_p$} \text{$|\Omega_d|^2$} - \text{$\Omega_s$} \text{$\Omega_c$} \text{$\Omega^{\ast}_d$}}{D},\nonumber \\
&&    \text{$\rho_{41}$}=
    \frac{\text{$\Omega_s$} \text{$|\Omega_c|^2$} - \text{$\Omega_p$} \text{$\Omega^{\ast}_c$} \text{$\Omega_d$}}{D},
    \label{Eq:rho_ij}
\end{eqnarray}
where $D=-\left[i \Gamma |\text{$\Omega_d$}|^2+(2\Delta +i \Gamma )|\text{$\Omega_c$}|^2\right]$. By substituting Eq.~(\ref{Eq:rho_ij}) into
MSEs [Eqs.~(\ref{Eq:MSEprobe}) and (\ref{Eq:MSEsignal})] with time-derivative components being zero, we obtain the steady-state solutions for
the probe and signal fields as follows:

\begin{eqnarray}
    \text{$\Omega_p(\alpha)$}=\frac{1}{\left|\Omega\right|^2}\left[\left|\text{$\Omega_c $}\right|^2\text{$\Omega_p$}(0)+\text{$\Omega_c$} \text{$\Omega_d^*$} \text{$\Omega_s$}(0)\right] \nonumber \\
    +\frac{1}{\left|\Omega\right|^2}\left[\left|\text{$\Omega_d$}\right|^2\text{$\Omega_p$}(0)-\text{$\Omega_c$} \text{$\Omega_d^*$} \text{$\Omega_s$}(0)\right]e^{-i\frac{\alpha }{2\xi }},
\label{Eq:probe}\\
    \text{$\Omega_s(\alpha)$}=\frac{1}{\left|\Omega\right|^2}\left[\left|\text{$\Omega_d $}\right|^2\text{$\Omega_s$}(0)+\text{$\Omega_d$} \text{$\Omega_c^*$} \text{$\Omega_p$}(0)\right] \nonumber
    \\
    +\frac{1}{\left|\Omega\right|^2}\left[\left|\text{$\Omega_c$}\right|^2\text{$\Omega_s$}(0)-\text{$\Omega_d$} \text{$\Omega_c^*$} \text{$\Omega_p$}(0)\right]e^{-i\frac{\alpha }{2\xi }},
\label{Eq:signal}
\end{eqnarray}
where $|\Omega|^2 = |\Omega_c|^2+|\Omega_d|^2$ and $\xi = i+2\frac{|\text{$\Omega_c$}|^2\Delta}{|\Omega |^2\Gamma}$. The terms $\Omega_p(0)$ and
$\Omega_s(0)$ represent the incident probe and signal fields, respectively. We then consider the phase of each laser field
$\Omega_j=|\Omega_j|e^{i\phi_j}$, where $j=p, s, c ,$ and $d$. The relative phase, $\phi_r$, of the four optical fields is defined as
$\phi_p-\phi_c+\phi_d-\phi_s$. Under the conditions $|\Omega_c|=|\Omega_d|$ and $|\Omega_p(0)|=|\Omega_s(0)|$, we obtain the simple steady-state
solutions for the probe and signal fields as follows:

\begin{eqnarray}
&&    \frac{\text{$\Omega_p(\alpha)$}}{\text{$\Omega_p(0)$}}=
    \frac{1}{2}\left[1+e^{-i\text{$\phi_r$}}+\left(1-e^{-i\text{$\phi_r$}}\right)e^{-i\frac{\alpha }{2\xi }}\right],
\label{Eq:probesimple}\\
&&    \frac{\text{$\Omega_s(\alpha)$}}{\text{$\Omega_s(0)$}}=
    \frac{1}{2}\left[1+e^{i\text{$\phi_r$}}+\left(1-e^{i\text{$\phi_r$}}\right)e^{-i\frac{\alpha }{2\xi }}\right].
\label{Eq:signalsimple}
\end{eqnarray}

The transmission and phase shift of the transmitted probe (signal) field are $|\Omega_{p(s)}(\alpha)/\Omega_{p(s)}(0)|^2$ and
tan$^{-1}$\{Im[$\Omega_{p(s)}(\alpha)$]/Re[$\Omega_{p(s)}(\alpha)$]\}, respectively. According to Eqs.~(\ref{Eq:probesimple}) and
(\ref{Eq:signalsimple}), when $\Delta = 0$ and $\phi_r=\pi$ the double-$\Lambda$ EIT medium becomes opaque and maximally attenuates both the
probe and signal fields. However, when $\phi_r=0$ both the probe and signal fields become completely transparent, as a result of destructive
interference. This phase-dependent double-$\Lambda$ EIT scheme with $\Delta = 0$ can be applied in low-light-level photon switching, as
previously described~\cite{ZhuDLambdaPS1}. Theories regarding the influence of the relative phase of the applied optical fields on the light
transmission when propagated through the double-$\Lambda$ medium were discussed in Ref.~\cite{KorsunskyDLambdaT}, and the matched propagation of
a pair of slow light pulses in the double-$\Lambda$ medium was studied in Ref.~\cite{DengDLambdaT}. Here, we focus on the effects of cross-phase
modulation (XPM) and coherent light amplification based on the double-$\Lambda$ EIT scheme.

%%%%%%%%%%%%%%%%%%%%%%%%%%%%%%%%%%%%%%%%%%%%%%%%%%%%%%%%%%%%%%%%%%%%%%%%%%%%%%%%
%%%%%%%%%%%%%%%%%%%%%%%%%%%%%%%%%%%%%%%%%%%%%%%%%%%%%%%%%%%%%%%%%%%%%%%%%%%%%%%%

\section{Results and discussions} \label{Sec:results}

To describe the mechanism and behavior of light pulses propagating in the double-$\Lambda$ EIT medium, the influence of the signal detuning
($\Delta$), the optical depth ($\alpha$), and the relative phase ($\phi_r$) for the optical medium are discussed in Sections~\ref{Sec:delta0}
and \ref{Sec:delta}. The initial phases of the probe and signal fields ($\phi_p$ and $\phi_s$) are set to 0 in the following calculations. We
draw a phase diagram to show the evolution of the phase shifts and transmission of both the probe and signal fields. The phase diagram reveals
that a clear phase jump that occurs when the relative phase is varied. We discuss the phase jump in Section~\ref{Sec:phasejump}. For practical
applications in optical and quantum control, an efficient XPM should satisfy a $\pi$-order phase shift with high transmission. We optimize the
parameters for achieving this goal and provide discussions in Section~\ref{Sec:transmission}. Finally, the effect of coherent light
amplification and the dynamics of both the probe and signal pulses propagating in the double-$\Lambda$ EIT medium are presented by numerically
simulating Eqs.~(\ref{Eq:OBEp41})--(\ref{Eq:MSEsignal}) in Section~\ref{Sec:slowlight}. In addition, we show that the steady-state transmission
obtained using the numerical simulations are consistent with those calculated using the analytical solutions [Eqs.~(\ref{Eq:probesimple}) and
(\ref{Eq:signalsimple})].

%%%%%%%%%%%%%%%%%%%%%%%%%%%%%%%%%%%%%%%%%%%%%%%%%%%%%%%%%%%%%%%%%%%%%%%%%%%%%%%%
\FigTwo
%%%%%%%%%%%%%%%%%%%%%%%%%%%%%%%%%%%%%%%%%%%%%%%%%%%%%%%%%%%%%%%%%%%%%%%%%%%%%%%%

\subsection{Balanced double-$\Lambda$ EIT system ($\Delta=0$)} \label{Sec:delta0}

We first discuss a symmetrical double-$\Lambda$ EIT system with a signal detuning of zero ($\Delta=0$). In the case where
$|\Omega_c|=|\Omega_d|$ and $|\Omega_p(0)|=|\Omega_s(0)|$, the two EIT systems are identical and the two FWM paths are balanced. Figures
\ref{fig:Fig2}(a) and \ref{fig:Fig2}(b) show phase diagrams of the probe and signal fields, respectively, plotted according to
Eqs.~(\ref{Eq:probesimple}) and (\ref{Eq:signalsimple}). In the phase diagram, the angle between the x-axis and a line connecting the origin and
data point represents the phase shift, and the square of the distance between the data point and the origin represents the transmission. To
illustrate the phase evolution of both the probe and signal fields propagating through the double-$\Lambda$ EIT medium, we increase the optical
depth, $\alpha$, from 0 to 100. The relative phase, $\phi_r$, is set from 1 to 6 as well as $\pi$, as shown in Figs. \ref{fig:Fig2}(a) and
\ref{fig:Fig2}(b). The dotted lines show the loops at various $\phi_r$ values and $\alpha=100$. Figures \ref{fig:Fig2}(c) and \ref{fig:Fig2}(d)
show the transmission and phase shifts of the probe (blue solid lines) and signal fields (red dashed lines). When $\Delta=0$, the two FWM
processes in the double-$\Lambda$ EIT system maintain a stable balance; hence, the variations of the probe and signal transmission according to
the relative phase are identical [see Fig. \ref{fig:Fig2}(c)]. Nevertheless, the signs of the variations in the phase shifts of the probe and
signal fields are opposite, as shown in Fig. \ref{fig:Fig2}(d). The phase shifts of the optical fields are continuous variations with $\phi_r$
and exhibit substantial changes in sign at $\phi_r=\pi$. When $\phi_r < \pi$, the probe (signal) filed acquires a negative (positive) phase
shift. By contrast, when $\phi_r > \pi$, the probe (signal) filed acquires a positive (negative) phase shift. However, when $\phi_r = \pi$, the
phase shifts of both the probe and signal fields are always zero.

%In addition, the evolutions of the fields terminate in the dotted lines for further increased optical density. The optical responses are
%insensitive to the optical density. {\color{red} Physical explanation or explain from Eqs.~\ref{Eq:probesimple} and \ref{Eq:signalsimple}.}

%%%%%%%%%%%%%%%%%%%%%%%%%%%%%%%%%%%%%%%%%%%%%%%%%%%%%%%%%%%%%%%%%%%%%%%%%%%%%%%%
\FigThree
%%%%%%%%%%%%%%%%%%%%%%%%%%%%%%%%%%%%%%%%%%%%%%%%%%%%%%%%%%%%%%%%%%%%%%%%%%%%%%%%

\subsection{Imbalanced double-$\Lambda$ EIT system ($\Delta\neq0$)} \label{Sec:delta}

The theoretical analysis reveals that this double-$\Lambda$ EIT scheme is phase dependent. Furthermore, a large phase shift (order $\pi$) of one
weak probe pulse induced by another weak signal pulse can be achieved, and hence this scheme can be applied in low-light-level XPM. However, as
shown in Figs.~\ref{fig:Fig2}(c) and \ref{fig:Fig2}(d), when $\phi_r$ is close to $\pi$, a large phase shift ($\approx \pi/2$) corresponds to
low transmission ($\approx e^{-\alpha}$), reducing practicality. Therefore, we consider a double-$\Lambda$ EIT system with a non-zero detuning
($\Delta\neq0$), which causes an imbalance between the two FWM processes of the double-$\Lambda$ EIT scheme.

We plotted phase diagrams with $\Delta =16.5 \Gamma$ and $\alpha$ ranging from 0 to 100, as shown in Fig.~\ref{fig:Fig3}. Based on the selected
parameters, a large phase shift with high light transmission can be generated, as discussed in Sec.~\ref{Sec:transmission}. The phase diagrams
show that the phase jump occurs in the probe field when $\phi_r=\phi_{pj}$ and in the signal field when $\phi_r=\phi_{sj}$. Here, we define
$\phi_{pj(sj)}$ as the relative phase when the phase jump occurs in the probe (signal) field. The curve of $\phi_r=\phi_{pj}$ or
$\phi_r=\phi_{sj}$ passing through the origin is a crucial condition for the phase jump [see Figs.~\ref{fig:Fig3}(a) and \ref{fig:Fig3}(b)].
When $\phi_r>\phi_{pj}$ (e.g., $\phi_r=5$ rad), the accumulated phase shift of the transmitted probe field becomes zero at approximately
$\alpha=40$, as indicated by the open circles in Fig.~\ref{fig:Fig3}(a). The probe field then exhibits a constantly increasing positive phase
shift until leaving the medium ($\alpha=100$). By contrast, when $\phi_r<\phi_{pj}$ (e.g., $\phi_r=4$ rad), the probe field constantly increases
negative phase shift throughout propagation. We conclude that the phase shift of the probe (signal) field as a function of $\phi_r$ must become
a phase jump when $\phi_r=\phi_{pj}$ ($\phi_r=\phi_{sj}$), as shown in Fig.~\ref{fig:Fig3}(d).

The double-$\Lambda$ EIT scheme in which $\Delta\neq0$ causes an imbalance between the two FWM paths and leads to that the probe and signal
fields exchange energy mutually. Hence, the energy flow causes the transmission of the probe or the signal field to be greater than unity [see
Fig.~\ref{fig:Fig3}(c)]. In addition, Figs.~\ref{fig:Fig3}(c) and \ref{fig:Fig3}(d) show large phase shifts (order $\pi$) with high light
transmission can be achieved by using the imbalanced double-$\Lambda$ EIT scheme.

%In addition, the large phase shift (order $\pi$) and the high light transmission increase the practicality of the scheme in optical control and
%communication [see Fig. 3(d)].
%the energy being transferred back and forth between two waves in each frequency pair, resonant and far detuned.

%%%%%%%%%%%%%%%%%%%%%%%%%%%%%%%%%%%%%%%%%%%%%%%%%%%%%%%%%%%%%%%%%%%%%%%%%%%%%%%%

\subsection{Phase jump} \label{Sec:phasejump}

%implying the light transmission becomes zero immediately during the propagation. With the further increased optical density, the light field
%revives. The distance between the point in the dotted line and the origin indicates the output transmission of the light field.

The key factor of the phase jump depends on whether the light field disappears during the propagation process (i.e., the curve in the phase
diagram passes through the origin) [see Figs.~\ref{fig:Fig3}(a) and \ref{fig:Fig3}(b)]. In this section, we determine the critical optical depth
($\alpha_c$) and relative phase ($\phi_{pj}$) when the phase jump occurs in the probe field. The curve in the phase diagram terminates at the
origin [i.e., $|\Omega_p(\alpha)/\Omega_p(0)|^2=0$ in Eq.~(\ref{Eq:probesimple})]; hence, we obtain the following equation:

\begin{eqnarray}
    \cot \left(\frac{\phi_{r}}{2}\right)e^{-R}+\tan \left(\frac{\phi_{r}}{2}\right)e^R = -2\sin (I),
\end{eqnarray}
where $R = -\frac{\alpha}{2}\frac{1}{(\Delta /\Gamma )^2+1}$ and $I = \frac{\alpha}{2}\frac{\Delta /\Gamma }{(\Delta /\Gamma )^2+1}$. We define
$\tan (\phi_{r}/2)e^R$ as $\chi$ and simplify the above expression as $\chi ^2+2\sin (I)\chi +1=0$. Hence, we derive

\begin{eqnarray}
    \chi = -\sin (I) \pm i\cos (I).
\end{eqnarray}
Because $\chi$ is a real number, $I=n\pi/2$, where $n$ is an odd integer. We then obtain the analytic solutions of $\alpha_c$ and $\phi_{pj}$ as
follows:

\begin{eqnarray}
&&    \alpha_c=n\pi\frac{(\Delta /\Gamma )^2+1}{\Delta /\Gamma }, \label{Eq:criticalOD}
\end{eqnarray}

\begin{eqnarray}
&&    \phi_{pj}=2 \tan ^{-1}\left[-\sin \left(\frac{n\pi}{2}\right)e^{\frac{n\pi}{2(\Delta /\Gamma )}}\right]. \label{Eq:phipj}
\end{eqnarray}
Similarly, the relative phase of the phase jump for the signal field can be derived from Eq.~(\ref{Eq:signalsimple}) as follows:
\begin{eqnarray}
    \phi_{sj}=2 \tan ^{-1}\left[\sin \left(\frac{n\pi}{2}\right)e^{\frac{n\pi}{2(\Delta /\Gamma )}}\right].
\label{Eq:phisj}
\end{eqnarray}

%%%%%%%%%%%%%%%%%%%%%%%%%%%%%%%%%%%%%%%%%%%%%%%%%%%%%%%%%%%%%%%%%%%%%%%%%%%%%%%%
\FigFour
%%%%%%%%%%%%%%%%%%%%%%%%%%%%%%%%%%%%%%%%%%%%%%%%%%%%%%%%%%%%%%%%%%%%%%%%%%%%%%%%

According to Eqs.~(\ref{Eq:criticalOD})-(\ref{Eq:phisj}) and using $\Delta=16.5\Gamma$ as an example, we obtain $\alpha_c\approx52$,
$\phi_{pj}\approx4.62$ rad, and $\phi_{sj}\approx1.66$ rad for $n=1$; these values are consistent with the numerical results shown in
Fig.~\ref{fig:Fig3}. The phase jump occurs when the light field disappears during the propagation process. As shown in the top plot of
Fig.~\ref{fig:Fig4}, the probe field exhausts its energy when the light propagates through a medium with a critical optical depth ($\alpha_c$),
and the signal field gains the energy. The system is converted into an EIT-based FWM system \cite{ChenFWM2014}. The probe field is then restored
when it passes through the remainder of the medium. We plotted the phase shifts of the probe (blue) and signal (red) fields with the relative
phase slightly above (solid lines) and below $\phi_{pj}$ (dashed lines), as shown in the bottom plot of Fig.~\ref{fig:Fig4}. The figure clearly
illustrates the phase jump near $\phi_{pj}$.

%%%%%%%%%%%%%%%%%%%%%%%%%%%%%%%%%%%%%%%%%%%%%%%%%%%%%%%%%%%%%%%%%%%%%%%%%%%%%%%%
\FigFive
%%%%%%%%%%%%%%%%%%%%%%%%%%%%%%%%%%%%%%%%%%%%%%%%%%%%%%%%%%%%%%%%%%%%%%%%%%%%%%%%

\subsection{Cross-phase modulation} \label{Sec:transmission}

For practicality in optical and quantum control, we discuss a $\pi$-order XPM with high light transmission achieved using the double-$\Lambda$
EIT scheme. To achieve a phase shift of $\pi$, the terminal point of the curve in the phase diagram must be located at the negative x-axis
[i.e., the imaginary part of Eq.~(\ref{Eq:probesimple}) is zero and the real part is negative]. We obtain the relative phase for the $\pi$-phase
shift of the probe field as follows:

\begin{eqnarray}
\phi_r^{\pi} =2 \tan^{-1}\left[\frac{\cos(I)-e^{-R}}{\sin(I)}\right]. \label{Eq:phipi}
\end{eqnarray}

In Eq.~(\ref{Eq:probesimple}), $\phi_r$ is substituted by Eq.~(\ref{Eq:phipi}) and we then obtain the light transmission as functions of
$\alpha$ and $\Delta$. We plotted the relationship between the probe field transmission and $\Delta$ with a fixed optical depth ($\alpha=100$)
in the inset of Fig.~\ref{fig:Fig5}(a). The gray zones in the figure show that the terminal point is located on the negative x-axis. The
transmission is maximal at approximately $\Delta=16.5 \Gamma$. Using various optical depths and the corresponding optimized $\Delta$, we obtain
a monotonous increasing function, as indicated by the black solid line in the main plot of Fig.~\ref{fig:Fig5}(a). A $\pi$ phase shift with high
light transmission can be achieved using the double-$\Lambda$ EIT scheme.

%%%%%%%%%%%%%%%%%%%%%%%%%%%%%%%%%%%%%%%%%%%%%%%%%%%%%%%%%%%%%%%%%%%%%%%%%%%%%%%%
\FigSix
%%%%%%%%%%%%%%%%%%%%%%%%%%%%%%%%%%%%%%%%%%%%%%%%%%%%%%%%%%%%%%%%%%%%%%%%%%%%%%%%

To apply the scheme in XPM, we compare the light transmission and phase shift with and without the signal field. In an ideal XPM technique, a
weak signal pulse can modulate another weak probe pulse by a phase shift of $\pi$ without losing energy. When no $\Omega_s$ is applied in the
proposed scheme, the probe transmission is a monotonous decreasing function of $\alpha$ with the corresponding $\Delta$, as indicated by the red
dashed line in Fig.~\ref{fig:Fig5}(a). Consider the parameters $\alpha=100$, $\Delta=16.5\Gamma$, and the corresponding $\phi_r$ for example.
Although the transmission can be as high as $68\%$ when $\Omega_s$ is present, the transmission becomes only $1\%$ when $\Omega_s$ is absent.
The phase modulation by the signal field, $|\Delta\phi_p^{\text{XPM}}|$, is 2.62 rad, as indicated by the black solid line in
Fig.~\ref{fig:Fig5}(c). Hence, achieving a $\pi$-order XPM with high light transmission by using the double-$\Lambda$ EIT scheme remains a
considerable challenge. In addition, we perform a similar simulation except the phase shift of the transmitted probe field, $|\Delta\phi_p|$, is
set to $\pi/2$. In this simulation, the real part of Eq.~(\ref{Eq:probesimple}) is zero and the terminal point in phase diagram is located on
the negative y-axis. As shown in Fig.~\ref{fig:Fig5}(b) and the blue dashed line in Fig.~\ref{fig:Fig5}(c), the probe transmissions are $140\%$
and $19\%$ with and without $\Omega_s$, respectively, and the $|\Delta\phi_p^{\text{XPM}}|$ is 0.57 rad when $\alpha=100$.

%Hence, to achieve near-$\pi$ XPM, high transmission cannot be achieved with and without the signal field by using the proposed scheme.

%%%%%%%%%%%%%%%%%%%%%%%%%%%%%%%%%%%%%%%%%%%%%%%%%%%%%%%%%%%%%%%%%%%%%%%%%%%%%%%%
%%%%%%%%%%%%%%%%%%%%%%%%%%%%%%%%%%%%%%%%%%%%%%%%%%%%%%%%%%%%%%%%%%%%%%%%%%%%%%%%

\subsection{Coherent light amplification} \label{Sec:slowlight}

The phase-dependent double-$\Lambda$ EIT system can coherently convert and amplify the energy of the light pulse. The two imbalanced FWM
processes cause two slow-light pulses (probe and signal) to exchange energy mutually and the light transmission can be greater than unity.
Numerical simulation of optimal energy amplification for the signal field (i.e., the highest signal transmission) versus optical depth with the
optimal detuning and relative phase reveals that an optical depth of 50 (100) enables attaining an amplification efficiency of approximately
84$\%$ (91$\%$), as indicated by the red dashed line in Fig.~\ref{fig:Fig6}(a).

%The scheme is useful for achieving high-efficiency frequency conversion.

The dynamics of pulse-shape light propagation by numerically solving MSEs and OBEs is presented in Figs.~\ref{fig:Fig6}(b) and
\ref{fig:Fig6}(c). Two identical square (probe and signal) pulses are fired into the double-$\Lambda$ medium simultaneously. To clearly observe
phase-dependent slow light pulses, we set $\alpha=100$, $\Delta=34.2\Gamma$, $\Omega_c=\Omega_d=1\Gamma$, and $\gamma_{21}=0$. The relative
phases $\phi_r$ of 1.53 rad and 4.76 rad enable achieving maximum probe (blue solid lines) and signal (red dashed lines) transmission, as shown
in Figs.~\ref{fig:Fig6}(b) and \ref{fig:Fig6}(c), respectively. The steady-state transmission is consistent with the results obtained by
calculating Eqs.~(\ref{Eq:probesimple}) and (\ref{Eq:signalsimple}). In addition, the group velocities differ between the transmitted probe and
the signal pulses because $\Delta\neq0$.
%%%%%%%%%%%%%%%%%%%%%%%%%%%%%%%%%%%%%%%%%%%%%%%%%%%%%%%%%%%%%%%%%%%%%%%%%%%%%%%%
%%%%%%%%%%%%%%%%%%%%%%%%%%%%%%%%%%%%%%%%%%%%%%%%%%%%%%%%%%%%%%%%%%%%%%%%%%%%%%%%

%%%%%%%%%%%%%%%%%%%%%%%%%%%%%%%%%%%%%%%%%%%%%%%%%%%%%%%%%%%%%%%%%%%%%%%%%%%%%%%%
%%%%%%%%%%%%%%%%%%%%%%%%%%%%%%%%%%%%%%%%%%%%%%%%%%%%%%%%%%%%%%%%%%%%%%%%%%%%%%%%

%%%%%%%%%%%%%%%%%%%%%%%%%%%%%%%%%%%%%%%%%%%%%%%%%%%%%%%%%%%%%%%%%%%%%%%%%%%%%%%%
%%%%%%%%%%%%%%%%%%%%%%%%%%%%%%%%%%%%%%%%%%%%%%%%%%%%%%%%%%%%%%%%%%%%%%%%%%%%%%%%

%\newpage

%%%%%%%%%%%%%%%%%%%%%%%%%%%%%%%%%%%%%%%%%%%%%%%%%%%%%%%%%%%%%%%%%%%%%%%%%%%%%%%%

%%%%%%%%%%%%%%%%%%%%%%%%%%%%%%%%%%%%%%%%%%%%%%%%%%%%%%%%%%%%%%%%%%%%%%%%%%%%%%%%